\documentclass[twocolumn,prb,showpacs]{revtex4}

\begin{document}
\topmargin-1.0cm

\title {Time-dependent current density functional theory for the linear
response of weakly disordered systems}
\author {C. A. Ullrich}
\affiliation {Department of Physics,  University of
Missouri-Rolla, Rolla, Missouri 65409}
\author {G. Vignale}
\affiliation {Department of Physics and Astronomy,  University of
Missouri-Columbia, Columbia, Missouri 65211}
\date{\today}

\begin{abstract}
Time-dependent density functional theory (TDFT) provides a way of calculating,
in principle exactly, the linear response of interacting many-electron systems,
and thus allows one to obtain their excitation
energies. For extended systems, there exist excitations of a collective
nature, such as bulk- and surface plasmons in metals or intersubband
plasmons in doped semiconductor quantum wells. This paper 
develops a quantitatively accurate first-principles description for
the frequency and the linewidth of such excitations in inhomogeneous
weakly disordered systems. A finite linewidth
in general has intrinsic and extrinsic sources. At low temperatures
and outside the region where electron-phonon interaction occurs, the only
intrinsic damping mechanism is provided by electron-electron interaction.
This kind of intrinsic damping can be described within TDFT, but one needs to
go beyond the adiabatic approximation and include retardation effects. It has
been shown [G. Vignale, C. A. Ullrich, and S. Conti, Phys. Rev.
Lett. {\bf 79}, 4878 (1997)] that a density-functional response theory that is
local in space but nonlocal in time has to be constructed in terms of the
currents, rather than the density. This theory will be reviewed in the first
part of this paper. For quantitatively accurate linewidths,
extrinsic dissipation mechanisms, such as impurities or disorder, have to be
included in the response theory. In the second part of this paper, we discuss
how extrinsic dissipation can be described within the so-called memory function
formalism. This formalisms will first be introduced and reviewed for
homogeneous systems. We will then present a synthesis of TDFT with the
memory function formalism for inhomogeneous systems, which allows one to
account simultaneously for intrinsic and extrinsic damping of collective
excitations.  As an example where both sources of dissipation are important
and where high-quality experimental data is available for comparison,
we discuss intersubband plasmons in a 40 nm wide GaAs/Al$_{0.3}$Ga$_{0.7}$As
quantum well.
\end{abstract}
\pacs{71.15.Mb; 71.45.Gm; 73.21.Fg; 78.67.De}
\maketitle
\newcommand{\xc}{{\rm xc}}
\newcommand{\qp}{q_{||}}
\newcommand{\pp}{p_{||}}
\newcommand{\lp}{l_{||}}
\newcommand{\kp}{k_{||}}
\newcommand{\yp}{y_{||}}
\newcommand{\qqp}{{\bf q}_{||}}
\newcommand{\ppp}{{\bf p}_{||}}
\newcommand{\rrp}{{\bf r}_{||}}
\newcommand{\llp}{{\bf l}_{||}}
\newcommand{\kkp}{{\bf k}_{||}}
\newcommand{\yyp}{{\bf y}_{||}}
\newcommand{\rp}{r_{||}}
\newcommand{\dy}{\displaystyle}
\newcommand{\ks}{_{\rm \scriptscriptstyle KS}}
\newcommand{\sinf}{\raisebox{-.7ex}{$\stackrel{<}{\sim}$}}
\newcommand{\ssup}{\raisebox{-.7ex}{$\stackrel{>}{\sim}$}}
\section{Introduction}
The calculation of excitation energies and linewidths of
collective excitations in extended electronic systems is one of
the outstanding problems in many-body theory. Time-dependent
density functional theory (TDFT)
\cite{RG,grosskohn,tdft,grodope,petersilka,leeuwen} offers a powerful and
elegant approach to this difficult problem. To set the stage for
the developments that are to follow, we shall begin this paper
with a summary of the key elements of TDFT. Let
\begin{equation} \label{H0}
\hat H_0 =  \sum_i \left\{ \frac{p_i^2}{2m} + v_{0}({\bf r}_i) \right\} +
\frac{1}{2} \sum_{i \neq j}U(|{\bf r}_i - {\bf r}_j|)
\end{equation}
be the Hamiltonian of a many-electron system, where ${\bf r}_i$ and ${\bf p}_i$
are the canonical coordinates and momenta of the $i$th electron, $m$ is its
mass, $v_{0}({\bf r})$ is a static external potential, which includes
contributions from randomly distributed impurities and other sources of 
disorder, and
$U(|{\bf r}_i - {\bf r}_j|)$ is the Coulomb interaction potential. To calculate
the excitation energies,\cite{grosskohn,petersilka} one adds to $\hat H_0$ a
small time-dependent perturbation of the form
\begin{equation}
\label{H1}
\hat H_1(t) =  \int\! d^3 r\:  v_1({\bf r},t) \hat n({\bf r}) \;,
\end{equation}
where $v_1({\bf r},t) = v_1({\bf r},\omega)e^{-i \omega t} + c.c.$
is a periodic potential ($v_1\ll v_0$) that couples linearly to
the density operator $\hat n({\bf r}) = \sum_i \delta({\bf r} -
{\bf r}_i)$. One then computes the time-dependent density of the
system, which, in the linear approximation, will be given by
\begin{equation} \label{N1}
n({\bf r},t) = n_0({\bf r}) + n_1({\bf r},\omega) e^{-i \omega t}+c.c. \;,
\end{equation}
where $n_0({\bf r})$ is the ground-state density, and $n_1({\bf r},\omega)$
is linearly related to $v_1({\bf r},\omega)$ via
\begin{equation}
\label{chi}
n_1({\bf r},\omega) = \int\!d^3r\: \chi({\bf r},{\bf r}',\omega)
v_1({\bf r}',\omega) \;.
\end{equation}
The {\it density-density response function} $\chi({\bf r},{\bf
r}',\omega)$ contains the essential information about those
excited states of the system that are coupled to the ground state
by the perturbation $\hat H_1(t)$. More specifically, in a finite
system (atom or molecule) this response function has a discrete
set of poles on the real frequency axis, corresponding to the
discrete excitation energies of the system. In an extended system,
the poles merge into a continuous branch cut along the real axis.
However, isolated poles can arise in the lower half of the complex
frequency plane: they correspond to collective excitations of the
system, where the imaginary part of the frequency defines the
characteristic lifetime of the excitation.

The basic TDFT strategy for calculating $\chi({\bf r},{\bf r}',\omega)$ is to
construct a {\it noninteracting} system that has the same ground-state density
$n_0({\bf r})$, and yields the same density response $n_1({\bf r},\omega)$ as
the interacting system under study. The dynamics of this noninteracting system
is controlled by an effective single-particle potential which is written as
the sum of the total external potential
$v_0({\bf r}) + v_1({\bf r},t)$ plus the Hartree potential
\begin{equation}
\label{hartree}
v_{\rm H}({\bf r},t) = e^2 \int\! d^3r'\:
\frac{n({\bf r}',t)}{|{\bf r} - {\bf r}'|}
\end{equation}
plus a remainder, which is known as the ``exchange-correlation'' (xc)
potential $v_{\xc}({\bf r},t)$. It is not at all obvious that such a potential
$v_{\xc}$ can be constructed, but, if it can, then Runge and Gross\cite{RG}
showed that it is a {\em unique functional} of the time-dependent density
up to within an additive function of time. The form of the xc potential
depends, in general, on the initial state of the system, but this
dependence disappears if one assumes, as we do here, that the system is
initially in its ground state. \cite{footnote} The effective noninteracting Hamiltonian
(also known as the Kohn-Sham Hamiltonian) that yields the exact density is
then given by
\begin{eqnarray}
\label{HKS}
\lefteqn{
\hat H\ks(t) = \sum_i \left\{ \frac{p_i^2}{2m} + v_{0}({\bf r}_i)+
v_{\rm H,0}({\bf r}_i)+v_{\xc,0}({\bf r}_i) \right\}} \nonumber \\
&+& \int\! d^3r \: \left[v_1({\bf r},t)+v_{\rm H,1}({\bf r},t)+v_{\xc,1}
({\bf r},t) \right ] \hat n({\bf r}) \;,
 \end{eqnarray}
where both the Hartree and the xc potential have been written as the sum of
static parts $v_{\rm H,0},v_{\xc,0}$ associated with the ground-state density,
and (small) time-dependent parts $v_{\rm H,1},v_{\xc,1}$ associated with the
time-dependent density. The static part of the Kohn-Sham Hamiltonian [first
line of Eq. (\ref{HKS})] yields the exact ground-state density in the
interacting system, while the time-dependent part [the second line of Eq.
(\ref{HKS})] yields the exact density response.

In  the linear response regime the xc potential can be written as
\begin{equation} \label{vxc}
v_{\xc}({\bf r},t) = v_{\xc,0}({\bf r}) + \int_{-\infty}^t
\!dt'\!\! \int\! d^3 r' \:
f_{\xc}({\bf r},{\bf r}',t-t')n_1({\bf r}',t') ,
\end{equation}
where $v_{\xc,0}({\bf r})$ depends only on the ground-state density, and
$f_{\xc}({\bf r},{\bf r}',t-t')$ is the retarded xc
kernel, formally defined as
\begin{equation}
f_{\xc}({\bf r},{\bf r}',t-t') = \left. \frac{\delta v_{\xc}[n]({\bf r},t)}
{\delta n({\bf r},t')} \right|_{n_0({\bf r})} .
\end{equation}
Fourier transformation of both $v_{\xc}$ and $f_{\xc}$ with respect to time
leads to the simpler relation
\begin{equation} \label{vxc1}
v_{\xc,1}({\bf r},\omega) =  \int\! d^3r'\: f_{\xc}({\bf r},{\bf r}',\omega)
n_1({\bf r}',\omega) \;.
\end{equation}
We denote by $\chi\ks({\bf r},{\bf r}',\omega)$ the
density-density response function of the {\it static} Kohn-Sham system
[the first line of Eq. (\ref{HKS})]. The second line of the same equation can
then be regarded as a time-dependent perturbation to the first. The
time-dependent density response is therefore given by
\begin{widetext}
\begin{equation} \label{n1bis}
n_1({\bf r},\omega) = \int\! d^3r'\:\chi\ks({\bf r},{\bf r}',\omega)
\left \{v_1({\bf r}',\omega) + \int \! d^3r''\left [{e^2 \over |{\bf r}'-
{\bf r}''|} + f_{\xc}({\bf r}',{\bf r}'',\omega) \right]
n_1({\bf r}'',\omega) \right \},
\end{equation}
where the second and the third terms in the curly brackets arise from
$v_{\rm H,1}$ and $v_{\xc,1}$, respectively. Comparing Eqs. (\ref{chi}) and
(\ref{n1bis}) we obtain the following integral relation between the exact
density-density response function $\chi({\bf r},{\bf r}',\omega)$ and the
noninteracting response function $\chi\ks({\bf r},{\bf r}',\omega)$:
\begin{equation} \label{chiintegralequation}
\chi({\bf r},{\bf r}',\omega) = \chi\ks({\bf r},{\bf
r}',\omega) + \int\! d^3x \:\chi\ks({\bf r},{\bf x},\omega)
\int \! d^3y \left [{e^2 \over |{\bf x} - {\bf y}|} + f_{\xc}({\bf
x},{\bf y},\omega) \right]\chi({\bf y},{\bf r}',\omega) \;.
\end{equation}
\end{widetext}
This can also be written as
\begin{equation} \label{inverserelation}
\chi^{-1}({\bf r},{\bf r}',\omega) =  \chi\ks^{-1}({\bf r},{\bf r}',
\omega) -{e^2 \over |{\bf r} - {\bf r}'|} - f_{\xc}({\bf r},{\bf r}',\omega),
\end{equation}
where $\chi^{-1}$ is the matrix inverse of $\chi$.

The excitation energies are finally obtained from the poles of the linear
response function, i.e., from the solution of the eigenvalue problem
\begin{equation}  \label{eigenvalueproblem}
\int \! d^3r'\:\chi^{-1}({\bf r},{\bf r}',\omega)  \Xi({\bf r}',\omega)= 0\;,
\end{equation}
where $\Xi({\bf r}, \omega)$ is the function that describes the spatial
dependence of the density in  the excited state.

Eq. (\ref{chiintegralequation}) is the main formal result in the
TDFT approach to the calculation of excitation energies. From a
fundamental point of view, calculating $f_{\xc}$ is of course
no easier than calculating $\chi$. The main advantage of recasting
linear response theory within TDFT is of a more practical nature:
as long as the exact excitations of the interacting system are in
qualitative correspondence to those of the Kohn-Sham system (a
kind of ``Fermi li\-quid'' assumption), the xc kernel in Eq.
(\ref{chiintegralequation}) is expected to be a small correction,
which can be approximated with relative impunity. \cite{footnote1}
The simplest approximation is to ignore both nonlocality in space
and retardation in time. This leads to the widely used {\it
adiabatic local density approximation} (ALDA)\cite{zangwill} in
which one poses
\begin{equation} \label{ALDA}
v_{\xc}({\bf r},t) = \left(\frac{d \epsilon_{\xc}(n)}{dn}
\right)_{n=n({\bf r},t)},
\end{equation}
where $\epsilon_{\xc}(n)$ is the xc energy density\cite{vosko} of the
homogeneous electron gas of density $n$. The right-hand side of Eq.
(\ref{ALDA}) is nothing but the local density approximation for the {\it
ground-state} xc potential evaluated at the time-dependent density. In terms
of the xc kernel, this approximation implies
\begin{equation}\label{fxc_alda}
f_{\xc}({\bf r},{\bf r}',\omega) =\frac{d^2 \epsilon_{\xc}(n)}{dn^2}\:
\delta ({\bf r} - {\bf r}') \;,
\end{equation}
which is a purely real and frequency-independent object.

The ALDA is a remarkably successful approximation, despite the
fact that it entirely neglects the frequency dependence of the xc
kernel, that is, the {\it retarded} dependence of the xc potential
on the density at earlier times. In atoms the ALDA has yielded
reasonably accurate values of the excitation energies.
\cite{petersilka,ogut} Most of the residual inaccuracy has been
traced to the fact that the ground-state xc potential in the LDA
fails at large distance from the nucleus. An optimized effective
potential approach,\cite{ulgogro} similar in spirit to Eq.
(\ref{ALDA}), but still without retardation, yields a dramatic
improvement in accuracy.\cite{grodope} Applications to more
complex systems (molecules,
polymers)\cite{baerends1,baerends2,baerends3} have met with
similar degree of success. The essential reason seems to be that
the frequency dependence of the xc kernel is rather weak, because
it is controlled by multi-electron excitations, which are either
very high in energy (atoms and molecules), or smoothly distributed
through a spectral range (extended systems).

There are, however, some important features of the dynamical response that
cannot be accounted for in any way by an instantaneous xc potential. Quite
generally, the need for a dynamical theory of $f_{\xc}$ arises in the study of
excitations that do not have an analogue in the Kohn-Sham system. Perhaps the
clearest example of this is provided by collective excitations in extended
electronic systems, such as bulk- and surface plasmons in metals or
inter- and intrasubband plasmons in doped semiconductor quantum
wells. In this case, the ALDA would predict resonance peaks of vanishing width,
in glaring contradiction to experiment.

Attempts to go beyond the ALDA to include retardation date back to the
mid-eighties. In 1985 Gross and Kohn\cite{grosskohn} proposed a dynamical
local density approximation for $f_{\xc}$ which was designed to preserve the
local relationship between $v_{\xc,1}$ and the density, while including
retardation in time. Their approximation reads
\begin{equation}
\label{GK}
f_{\xc}({\bf r},{\bf r}',\omega) = f^h_{\xc}(k=0,\omega)
\delta({\bf r} - {\bf r}') \;,
\end{equation}
where $f^h_{\xc}(k,\omega)$ is the xc kernel of the uniform electron gas
calculated at the local ground-state density $n_0({\bf r})$ (more about
$f^h_{\xc}$ will be said in the next sections). Because $f_{\xc}(\omega)$ is
complex, this approximation yields a finite linewidth for excitations that
would have zero linewidth in  the ALDA. \cite{dobson,footnote2}

Unfortunately, the Gross-Kohn approximation (\ref{GK}) suffers from several
inconsistencies, such as the failure to satisfy the generalized Kohn's theorem
\cite{kohntheorem} and related sum rules.\cite{dobson1,VK} As a consequence,
it was found \cite{ullrichvignale} that within this approximation,
intersubband plasmons in quantum wells may become substantially overdamped.
These deficiencies were ultimately
traced back to the fact that a local approximation for the dynamical xc
potential in terms of the density does not exist (except at $\omega =0$,
in which case it is the static LDA). The reason for this startling result is
that the xc kernel of a non-homogeneous system is a function of infinite
range in space, or, more precisely, the spatial Fourier transform
$f_{\xc}({\bf k}, {\bf k}',\omega)$ diverges when $k \to 0$ at constant
$k'$ or vice versa. (In the homogeneous case one has ${\bf k}={\bf k}'$
and the singularity disappears).

Vignale and Kohn (VK) \cite{VK} and Vignale, Ullrich and Conti
(VUC) \cite{VUC} showed that the nonlocality problem could be
circumvented by working with the {\em current density} rather than
the density as a basic variable. The idea is to perturb the system
(\ref{H0}) with a time-dependent {\it vector potential} ${\bf
a}_1({\bf r},t) = {\bf a}_1({\bf r},\omega)e^{-i \omega t}+c.c.$
rather than with a scalar potential. The perturbing Hamiltonian
has the form
\begin{equation}
\hat H_1(t) =  \int\!d^3r\: {\bf a}_1({\bf r},t) \cdot \hat{\bf j}_p
({\bf r}) \;,
\end{equation}
where $\hat{\bf j}_p({\bf r}) = {1 \over 2m}\sum_i \left [ \hat {\bf p}_i
\delta ({\bf r} - {\bf r}_i) + \delta ({\bf r} - {\bf r}_i) \hat {\bf p}_i \right ]$ 
is the {\it paramagnetic} current density operator. \cite{footnote4}
One then calculates the current response, and determines the
excitation energies from the poles of the {\it current-current response
function}.

The Kohn-Sham Hamiltonian in this time-dependent current density functional
theory (TCDFT)  contains an xc vector potential ${\bf a}_{\xc,1}
({\bf r},t)$, which is a (linear) functional of the full current density
response
${\bf j}_1({\bf r},t)= {\bf j}_{p1}({\bf r},t)+n_0({\bf r}){\bf
a}_1({\bf r},t)/m$:
\begin{equation} \label{vxc1bis}
a_{\xc,1,\alpha}({\bf r},\omega) = \sum_\beta\! \int d^3r'\:
f_{\xc,\alpha\beta}({\bf r},{\bf r}', \omega) j_{1,\beta}({\bf r}',\omega),
\end{equation}
where $f_{\xc,\alpha\beta}({\bf r},{\bf r}',\omega)$ is the tensorial
generalization of the usual xc kernel (here and in the following, $\alpha,\beta$
denote Cartesian components). The static part of the Kohn-Sham Hamiltonian
remains unchanged, and the ground-state density is still determined by the
static xc field $v_{\xc,0}$.

It turns out that the xc vector potential {\em does} admit a local
approximation in terms of the current density: as we shall see in
the next section, the form of this approximation is essentially
determined by symmetry considerations and can be expressed in
terms of an xc stress tensor.\cite{VUC} The resulting expression
for ${\bf a}_{\xc,1}$ is local in space, retarded in time,
satisfies the generalized Kohn's
theorem,\cite{kohntheorem,dobson1} and allows a consistent
calculation of the linewidth of elementary excitations, at least
the part of it that arises from intrinsic many-body effects. The
fundamental reason why all this is possible is that the
relationship between the longitudinal current and the density is
nonlocal. From the continuity equation
\begin{equation} \label{continuity}
\frac{\partial n_1({\bf r},t)}{\partial t}= - \nabla \cdot{\bf j}_1({\bf r},t)
\end{equation}
one sees that the longitudinal component of the current is given by
\begin{equation} \label{continuity2}
j_{1,L}({\bf r},t) = \frac{1}{4\pi}\int\! d^3r' \: \frac{n_1({\bf r}',t)}
{| {\bf r} - {\bf r}' |} \;,
\end{equation}
while the transverse component of the current remains undetermined. Thus,
a local functional of $j_{1,L}$ will necessarily be a nonlocal functional of the
density. What is remarkable here is that the nonlocality of
$v_{\xc,1}$ as a functional of density can be completely eliminated by
``upgrading''
TDFT to a description in terms of ${\bf a}_{\xc,1}$ and ${\bf j}_1$.
This is the essence of the VK and VUC theory.

Since any scalar potential can be represented by an equivalent vector
potential, and since the density is easily calculated from the current
[see Eq. (\ref{continuity})], we see that the ordinary TDFT is a special case
of the TCDFT formulation. \cite{footnote3} An additional advantage of this
formulation is that it allows one to treat the more general problem of the
response of an electronic system to an electromagnetic field having both
longitudinal and transverse components, whereas the original Runge-Gross
formulation is limited to longitudinal fields, i.e., fields that can be
expressed as the gradient of a scalar potential.

Although the VK and VUC formulations are important steps enabling
the calculation of the linewidth of elementary excitations in
extended systems, they are still not sufficient to achieve
quantitative accuracy in cases of practical interest. For example,
the calculation of the linewidth of the intersubband (ISB) plasmon
in a 40-nm GaAs/Al$_{0.3}$Ga$_{0.7}$As quantum well reported in
Ref. \onlinecite{ullrichvignale}, based on VUC formalism, yielded
a linewidth about 5 times smaller than the experimental value.
The reason for this disappointing result is that the theory, as it
stands, does not take into account other intrinsic and extrinsic sources of damping,
such as electron-phonon interactions, electron-impurity scattering, 
and, in the case of quantum wells, interfacial roughness. All
these interactions contribute to the linewidth and must be
included in any calculation that aspires to quantitative accuracy.

In this paper we take a first step in this direction by showing how two of
the most prominent contributions to the low-temperature linewidth of plasmons
in quantum wells, namely electron-impurity scattering and interfacial
roughness, can be built into the current-density functional formalism.

Our approach is based on a the ``marriage'' of the TCDFT formalism with the
memory function formalism described, for example, by Forster.\cite{forster}
In the homogeneous electron gas limit this approach reduces to the
Belitz-Das Sarma \cite{belitz} treatment of the effect of impurities on bulk
plasmons, which, in turn, can be viewed as the high-frequency extension of the
Mermin relaxation-time approximation \cite{mermin}
for the density-density response function of an electron gas
in the presence of randomly distributed impurities. Of course, our interest
lies in strongly inhomogeneous systems, such as quantum wells,
\cite{prl,jon} which exhibit
the {\it intersubband} plasmon resonance. Such resonances are of practical
interest  in connection with the design of infrared detector devices.

Our strategy is to derive an integral equation which relates the current
response function of the disordered interacting many-electron system to that
of the same system in the absence of disorder: the latter is calculated by
the standard TCDFT outlined above. We shall show that this approach (despite
some inevitable approximations in the treatment of disorder) meets with
considerable success: the linewidth of the intersubband plasmon is considerably
enhanced by disorder -- in particular, by interfacial roughness -- and
agrees quantitatively with the measured one. More importantly, the qualitative
behavior of the linewidth as a function of an external electric field that
controls the shape of the quantum well is correctly reproduced.

The remainder of this paper is organized as follows: In Section
\ref{secII} we review the main aspects of the TCDFT formalism for
the linear response of many-electron systems. In Section
\ref{secIII} we review the memory function formalism and
demonstrate its application to the case of a homogeneous electron
gas with randomly distributed impurities. These sections are meant
to make the paper self-contained and to provide the necessary
background for the following more technical parts. In Section
\ref{secIV} we combine the memory function formalism with TCDFT
for inhomogeneous systems and derive the key integral equations
for the current-current response functions. Finally, in Section
\ref{secV} we demonstrate the power of the method by calculating
the linewidth of the intersubband plasmon in a quantum well and
comparing to recent experimental results, and in Section
\ref{secVI} we give our conclusions.

\section{TCDFT beyond the adiabatic LDA } \label{secII}
\subsection {Exchange-correlation kernels in the homogeneous electron gas}

For orientation, let us first consider the xc kernels of a
{\it homogeneous } electron gas. Because of translational invariance, it is
convenient to work with the Fourier transform ${\bf j}({\bf k},\omega)$ of the
current density. The linear response of this quantity to a vector potential
${\bf a}_1({\bf k},\omega)$ can be written as
\begin{equation} \label{currentresponse1}
j_{1,\alpha}({\bf k},\omega) = \sum_\beta \chi_{\alpha\beta}({\bf k}, \omega)
a_{1,\beta}({\bf k},\omega) \;,
\end{equation}
where $\chi_{\alpha\beta}({\bf k},\omega)$ is the current-current response
tensor.
Due to rotational invariance, the responses of the longitudinal (parallel
to ${\bf k}$) and transverse (perpendicular to ${\bf k}$) components of the
current are completely independent, and one can write
\begin{equation} \label{currentresponse2}
j_{1,L(T)}({\bf k},\omega) = \chi_{L(T)}({\bf k}, \omega)
a_{1,L(T)}({\bf k},\omega) \;,
\end{equation}
where $L(T)$ denotes the longitudinal (transverse) component, and
$\chi_{L(T)}({\bf k},\omega)$ is the longitudinal (transverse) response
function. According to the general linear response formalism,\cite{pinesnozieres}
$\chi_{L(T)}$ is given by
\begin{eqnarray} \label{currentresponse3}
\chi_{L(T)}({\bf k},\omega)  &=& \frac{n}{m}+\sum_l |\langle l |\hat j_{p,L(T)}
({\bf k}) |0\rangle |^2 \nonumber\\
&\times& \!\! \left \{ {1 \over \omega - \omega_{l0} + i\eta} -
{1 \over \omega + \omega_{l0} + i\eta} \right \},
\end{eqnarray}
where $|0\rangle$ is the ground state, $|l\rangle$ is the $l$-th excited state,
and $\omega_{l0}$ is the excitation energy $E_l-E_0$. The Fourier
transform of the paramagnetic current operator $\hat {\bf j}_p({\bf k})$ is
given by
\begin{equation} \label{paracurrent}
\hat {\bf j}_p({\bf k}) = \frac{1}{2m} \sum_i \left[
\hat {\bf p}_i e^{-i {\bf k} \cdot {\bf r}_i} +
e^{-i {\bf k} \cdot {\bf r}_i} \hat {\bf p}_i \right ].
\end{equation}
Note that the total current response is the sum of
the ``London current'' $n {\bf a}_1/m$ and the paramagnetic current [the
expectation
value of (\ref{paracurrent})]. A key feature of Eq.
(\ref{currentresponse3}) is that
in the limit of $k \to 0$ and finite $\omega$ it has the limiting form
\begin{eqnarray} \label{chismallk}
\chi_L(k,\omega) & \to & \frac{n}{m \epsilon(\omega)} +
\alpha_L(\omega)k^2 \nonumber \\
\chi_T(k, \omega) &\to& \frac{n}{m} + \alpha_T (\omega) k^2,
\end{eqnarray}
where $\alpha_{L(T)}(\omega)$ are functions of frequency only,
$\epsilon(\omega) = 1 - \lim_{k \to 0} n v(k) k^2 / m \omega^2$ is the
homogeneous dielectric function, and $v(k)$ is the Fourier transform of the
Coulomb interaction. Notice that the difference between the longitudinal and
transverse results at $k=0$ is due to the long range of the Coulomb
interaction: the difference vanishes if $v(k)$ diverges more slowly than
$1/k^2$.

Translational invariance is the essential reason for the small-$k$ behavior of
the $\chi$'s. In the $k \to 0$ limit the current operator reduces to the total
momentum operator plus a correction that vanishes linearly with $k$. Thus the
first term on the right-hand side of Eq. (\ref{chismallk}) is the response of
the center-of-mass momentum, which obeys a simple equation of motion under the
action of the external force, while the second term, of order $k^2$, comes
from the residual part of the operator, which is linear in $k$. There are no
cross terms, since the dynamics of the center of mass is decoupled from that
of the internal degrees of freedom.

Let us now turn to the xc potentials. The idea is to express the exact current
response (\ref{currentresponse2}) as the response of a noninteracting electron
gas to an effective vector potential, written as
\begin{equation}
{\bf a}_{\rm eff,1} ({\bf k}, \omega) = {\bf a}_1 ({\bf k},\omega) +
{\bf a}_{\rm H,1} ({\bf k},\omega) + {\bf a}_{\xc,1} ({\bf k},\omega) \;.
\end{equation}
The Hartree component ${\bf a}_{\rm H,1}$ is purely longitudinal (since it is
just another way of describing the scalar Hartree potential) and is given by
\begin{equation} \label{Ahartree}
{\bf a}_{\rm H,1}({\bf k},\omega) = \frac{k^2}{\omega^2} \: v(k)
j_{1,L}({\bf k},\omega) \hat {\bf k}\;.
\end{equation}
The xc potential can be decomposed into its longitudinal and transverse
components (with respect to the direction of ${\bf k}$) as follows:
\begin{eqnarray} \label{axc}
{\bf a}_{\xc,1}({\bf k},\omega) &=& \frac{k^2}{\omega^2} \left[
f^h_{\xc,L}(k,\omega) j_{1,L}({\bf k},\omega) \hat {\bf k} \right. \nonumber\\
&+& \left.
f^h_{\xc,T}(k,\omega) {\bf j}_{1,T}({\bf k}, \omega) \right].
\end{eqnarray}
The factor $k^2 /\omega^2$ has been introduced, in analogy to (\ref{Ahartree}),
so that the longitudinal component of the xc vector potential is equivalent to
the scalar xc potential
\begin{equation}
v_{\xc,1}({\bf k},\omega) = \frac{\omega a_{\xc,1,L}}{k} =
f^h_{\xc,L}(k,\omega) n_1({\bf k}, \omega)
\end{equation}
($j_L=n_1 \omega/k$), thus making $f^h_{\xc,L}(k,\omega)$ identical with
the usual $f^h_{\xc}(k,\omega)$ of the ordinary density functional theory.\cite{grosskohn}

In analogy to Eq.~(\ref{inverserelation}), the relationship between the
interacting current-current response function and its noninteracting counterpart
$\chi\ks$ takes the form
\begin{eqnarray} \label{chim1}
\chi^{-1}_{L(T)}(k,\omega) &=& \chi^{-1}_{{\rm \scriptscriptstyle KS},L(T)}(k,\omega)
\nonumber\\
&-& {k^2
\over \omega^2} \left [v_{L(T)}(k) + f^h_{\xc,L(T)}(k,\omega) \right ],
\end{eqnarray}
where we have defined $v_L(k) = v(k)$ and $v_T(k)=0$.

Notice that, according to Eqs.~(\ref{chismallk}), $\chi^{-1}_L(k,\omega)$ and
$\chi^{-1}_{{\scriptscriptstyle \rm KS},L}(k,\omega) - k^2v_L(k)/\omega^2$ have the same limit $m
\epsilon(\omega) /n +O(k^2)$ for $k \to 0$ and finite $\omega$.   Similarly,
$\chi^{-1}_T(k,\omega)$ and
$\chi^{-1}_{{\rm \scriptscriptstyle KS},T}(k,\omega) - k^2v_T(k)/\omega^2$ have the same limit $m/n
+O(k^2)$.  Thus, we see that Eq.~(\ref{chim1}) is consistent with the
limiting forms
(\ref{chismallk}) if and only if   the $k \to 0$ limits of the xc kernels
$f^h_{\xc,L(T)}$
are finite functions of frequency:
\begin{equation} \lim_{k \to 0} f^h_{\xc,L(T)}(k,\omega) \equiv
f^h_{\xc,L(T)}(\omega).
\end{equation}
Because of the central role  these functions play in the developments to
follow,
we now  describe their properties in detail.

\subsection{Properties of the homogeneous xc kernels}

The calculation of the xc kernels $f^h_{\xc,L(T)} (\omega)$ is a very
difficult problem in many-body theory. Approximate calculations
have been done using (i) Interpolation schemes between exact high
and low frequency limits \cite{grosskohn} (ii) Perturbation theory
\cite{glicklong,holassingwi} (iii) Mode decoupling approximations.
\cite{nifosi} Here we simply summarize the main results that have
been established to date, and refer the reader to the original
references.

1. The high-frequency limit is a purely real constant  given by
\begin{equation}
\lim_{\omega \to \infty} f^h_{\xc,L(T)} (\omega) = {1 \over 2n}
\! \left [ d_{L(T)}
(\langle ke\rangle -\langle ke\rangle _0 ) + e_{L(T)} \langle pe\rangle
\right ],
\end{equation}
which is also known as the {\it third moment sum rule}. $\langle ke
\rangle$ and $\langle pe \rangle$ are the expectation values of the
kinetic and potential energy, respectively, and $\langle ke \rangle_0$ is 
the noninteracting kinetic energy. In three dimensions, $d_L=4$,
$e_L=8/15$, $d_T=4/3$,
$e_T=-4/15$. In two dimensions,  $d_L=6$, $e_L=5/4$, $d_T=2$, $e_T=-1/4$
(see Ref. \onlinecite{nifosi}).

The behavior of the imaginary part of the longitudinal xc kernel
was first determined by Glick and Long \cite{glicklong} (3D) and
Holas and Singwi \cite{holassingwi} (2D), making use of
second-order perturbation theory, which becomes, in all likelihood, exact in
the high-frequency limit. More recently, their calculation has
been confirmed and extended to the transverse kernel \cite{nifosi}
by a different method based on the equations of motion for the
current response function.  The result is
\begin{equation} \label{fxc_limit}
\lim_{\omega \to \infty}  {\rm Im} f^h_{{\rm xc},L(T)}(\omega) = - a_{L,T}\:
\pi^{4-D} \bar \omega^{-D/2} {e^2 \over a_0^{1-D}} \;,
\end{equation}
where $a_0 = \hbar^2/me^2$ is the Bohr radius,  $\bar \omega =
\hbar \omega a_0/e^2$ is the dimensionless frequency, and $D$ is the number of
spatial dimensions.  The coefficients are $a_L=23/30$ and $a_T=16/15$ in three
dimensions, and $a_L=11/32$ and $a_T=9/32$ in  two dimensions.

2. The $\omega \to 0$ limit
was first worked out by Conti and Vignale,\cite{contivignale}
and is subtly different from the static limit, which is obtained by setting
$\omega = 0$ {\it before}  letting $k \to 0$. The result for $f_{{\rm xc},L}$  in $D$
dimensions is
\begin{equation} \label{zerolimit}
\lim_{\omega \to 0}  f^h_{\xc,L}(\omega) =  \epsilon''_{\xc}(n) \ + {2(D-1)
\over D}
\lim_{\omega \to 0} f^h_{\xc,T} (\omega)  ,
\end{equation}
where $\epsilon''_{\xc}(n) \equiv d^2 \epsilon_{\xc}(n)/dn^2$.
The second term on the right-hand side of this expression is proportional to
\begin{equation}
\lim_{\omega \to 0} f^h_{\xc,T}(\omega) = {2 E_F \over 25 n}\left({3F_2 - 5F_1
\over 3+F_1}\right),
\end{equation}
in three dimensions, and
\begin{equation}
\lim_{\omega \to 0} f^h_{\xc,T}(\omega) = {E_F \over 2 n}\left({F_2 - F_1  \over 2
+F_1}\right)
\end{equation}
in two dimensions, where $F_l$ are the conventional Landau parameters 
of the electron liquid. \cite{pinesnozieres}
Note that the first term on the right-hand side of  Eq. (\ref{zerolimit})
is the usual compressibility  obtained from the static limit $\lim_{k \to 0}
\lim_{\omega \to
0}$ $f^h_{\xc,L} (k,\omega)$.  Thus,  Eq. (\ref{zerolimit})  vividly shows the
non-commutativity of the $k \to 0$ and $\omega \to 0$ limits.

3.
Due to the causality properties of the linear response functions, the xc
kernels must be
analytic functions of $\omega$ in the upper half of the complex plane. This
leads to
the Kramers-Kronig relations, which relate the real parts of the xc kernels
to their
imaginary parts:
 \begin{equation} \label{KK}
{\rm Re} f^h_{\xc}(\omega) = f^h_{\xc}(\infty) +  \frac{1}{\pi}
P \int\! d\omega' \: { {\rm Im} f^h_{\xc}(\omega ')
\over \omega' - \omega } \; ,
\end{equation}
where $P$ denotes  the ``principal part" integral.

4. In a very recent development \cite{zhixinvignale} the
low-frequency behavior of the imaginary parts of the xc kernels has also been
calculated exactly to leading order in the strength of the Coulomb interaction.
The results are
\begin{equation}\label{imzerolimit1}
{\rm Im} f_{{\rm xc},T} (\omega) \to - {1 \over (n a_0^D)^2} \:  {\eta
\over (\hbar/ a_0^D)}  \bar \omega \:
{e ^2 \over a_0^{1-D}},
\end{equation}
and
\begin{equation}\label{imzerolimit2}
{\rm Im} f_{{\rm xc},L} (\omega) \to {2(D-1) \over D} \: {\rm Im} f_{xc,T} (\omega),
\end{equation}
 where the dimensionless ``shear viscosity"  $ {\eta \over (\hbar/ a_0^D)}$ is given
by
\begin{eqnarray}
  {\eta \over (\hbar/ a_0^D)} &=& - \frac{k_Fa_0} {45 \pi^3} 
  \biggl \{ 5 -
\left(\lambda +\frac{5}{\lambda}\right) \tan^{-1} \lambda \nonumber\\
&&-\frac{2}{\lambda} \sin^{-1} \frac{\lambda}{\sqrt{1 + \lambda^2 }}  \nonumber\\
&&+ \frac{2}{\lambda \sqrt{2 +\lambda^2}} \left[ \frac{\pi}{2}
-\tan^{-1} \frac{1}{\lambda \sqrt{2 + \lambda^2}} \right] \biggr \}
\end{eqnarray}
in three dimensions, and
\begin{eqnarray} \label{slope2d}
{\eta \over (\hbar/ a_0^D)} &=&\frac{1}{12 \pi^2} \biggl \{ 2
\left[\ln(\lambda +1)
-\frac{\lambda}{1+\lambda} \right]   \nonumber\\
&& -\int_0^1 dx \frac{\lambda^2 x}
{(\lambda x +1)(\lambda \sqrt{1- x^2} +1)} \biggr \}
\end{eqnarray}
in two dimensions. In the above expressions $\lambda$  is defined as
$\lambda = 2 k_F /k_s$, where $k_s$ is the screening wavevector, e.g., in
RPA, $k_s = (4 \pi   k_Fa_0)^{1/2}/\pi a_0$ in three dimensions, and $k_s = 2/a_0$
in two dimensions. The  derivation of these results is presented in Ref.
\onlinecite{zhixinvignale}.

5.  Parametrized expressions.
To keep our presentation self-contained, we   also include the explicit
parametrization for $f_{\xc, L}(\omega)$  that has been used in the
calculations of Section V.  This is the original Iwamoto-Gross-Kohn 
parametrization, 
\cite{grosskohn,iwamoto} and has the form
\begin{equation}
{\rm Im} f_{\xc, L}(\omega) = {a(n) \omega \over [1 + b(n )\omega^2]^{5/4}}
\end{equation}
with the coefficients $a(n)$ and $b(n)$ determined by the compressibility
and third
moment sum rules, and the Kramers-Kronig dispersion relations.  The real part of
$f_{\xc, L}(\omega)
$ is then calculated with the help of the dispersion relation (\ref{KK}).
More recent analytic expressions for  $f_{\xc ,L}(\omega)$  and $f_{\xc,
T}(\omega)$  have been obtained by Nifosi, Conti and Tosi \cite{nifosi}
and Qian and Vignale (QV). \cite{zhixinvignale}
These new expressions possess considerable structure in the
frequency dependence due to two-plasmon excitations.  The QV expression reproduces the
exact
perturbative limit of   $f_{\xc }(\omega)$  in the limit $\omega \to 0$.
Additional details about these expressions can be found in the original references.

\begin{widetext}
%

\subsection{The exchange-correlation field for a homogeneous electron gas}
As a preparation for the study of inhomogeneous systems let us now
examine the real-space form of the xc vector potential. It is
convenient for this purpose to introduce the xc electric field
\begin{eqnarray}
\label{exc}
{\bf E}_{\xc,1}({\bf k},\omega) &=& i \omega {\bf a}_{\xc,1}({\bf k},\omega)
\nonumber \\
&=&-\frac{1}{i \omega} \Big\{ {\bf k} [{\bf k} \cdot {\bf j}_1({\bf k},\omega)]
f^h_{\xc,L}(\omega) + k^2 f^h_{\xc,T}(\omega) {\bf j}_{1,T} ({\bf k},\omega)
\Big\}.
\end{eqnarray}
Splitting off the familiar ALDA contribution
\begin{equation} \label{exclda}
{\bf E}_{\xc,1}^{\rm ALDA}({\bf k},\omega) = i {\bf k} \epsilon_{\xc}''
{{\bf k} \cdot {\bf j}_1 \over \omega} \;,
\end{equation}
we can write
\begin{eqnarray} \label{exc2}
{\bf E}_{\xc,1}({\bf k},\omega) &=& {\bf E}_{\xc,1}^{\rm ALDA}({\bf k},\omega)
\nonumber \\&-&
\frac{1}{i \omega} \Big\{ {\bf k}[{\bf k} \cdot {\bf j}_1({\bf k},\omega)]
[f^h_{\xc,L}(\omega)-\epsilon_{\xc}'']
+ k^2 f^h_{\xc,T}(\omega) {\bf j}_{1,T} ({\bf k}, \omega) \Big\}.
\end{eqnarray}
Introducing at this point the velocity field
\begin{equation}
{\bf u} = \frac{{\bf j}_1}{n}
\end{equation}
and Fourier transforming Eq. (\ref{exc2}) to real space, we get
\begin{eqnarray} \label{exc3}
{\bf E}_{\xc,1}({\bf r},\omega) &=& {\bf E}_{\xc,1}^{\rm ALDA}({\bf r},\omega)
\nonumber \\&+&
\frac{n}{i\omega} \Big\{ \left [f^h_{\xc,L}(\omega)  -  f^h_{\xc,T}(\omega) -
\epsilon_{\xc}^{\prime \prime} \right ]
\nabla( \nabla \cdot {\bf u} ) + f^h_{\xc,T}(\omega) \nabla^2 {\bf u}\Big\}.
\end{eqnarray}
It is easy to verify that  this expression can be re-written as
\begin{equation} \label{exc5}
E_{\xc,1,\alpha}({\bf r},\omega) = E_{\xc,1,\alpha}^{\rm ALDA}({\bf r},\omega)+
{1 \over n} \sum_\beta \frac{\partial \sigma_{\xc,\alpha\beta}({\bf r},
\omega)}
{\partial r_\beta}  \;,
\end{equation}
where the xc stress tensor  $\sigma_{\xc,\alpha \beta}$ is defined as
  \begin{equation} \label{sigmaxc}
\sigma_{\xc,\alpha\beta} = \tilde \eta (n,\omega) \left(
{\partial u_\alpha \over \partial r_\beta}+ {\partial u_\beta \over
\partial r_\alpha}-  {2 \over D} \nabla \cdot {\bf u}\,
\delta_{\alpha\beta} \right )
+ \tilde \zeta (n,\omega) \nabla \cdot {\bf u}\, \delta_{\alpha\beta} \;,
\end{equation}
\end{widetext}
and
\begin{equation} \label{visco_eta}
\tilde \eta (n,\omega) = -{n^2 \over i \omega}f^h_{\xc,T}(\omega)
\end{equation}
\begin{equation}
 \tilde \zeta (n,\omega) = -{n^2 \over i \omega} \left(
f^h_{\xc,L}(\omega) - {2(D-1) \over D} f^h_{\xc,T}(\omega) -
\epsilon_{\xc}^{\prime \prime} \right ) \label{visco_zeta}
\end{equation}
are generalized (i.e., frequency-dependent and complex) visco-elastic
constants of the electron liquid.\cite{contivignale}
In particular, the {\it real} parts of  $\tilde \eta$ and $\tilde \zeta$
[related to the imaginary parts of $f^h_{\xc,L(T)}(\omega)/\omega$] play the role of
shear and bulk viscosities, respectively, while the imaginary parts of
$\omega \tilde \eta$ and  $\omega \tilde \zeta$ [related to the real parts of
$f^h_{\xc,L(T)}(\omega)$] are interpreted as post-ALDA  xc
contributions to frequency-dependent elastic constants $\mu$ (shear modulus) and
$K$ (bulk modulus) of the electron liquid: $\mu_{\rm dyn} = n^2 {\rm Re}
f^h_{\xc,T}(\omega)$ and $K_{\rm dyn} = n^2 {\rm Re}\{ f^h_{\xc,T}(\omega) -
[2(D-1)/D] f^h_{\xc,T}(\omega) - \epsilon_{\xc}^{\prime \prime} \}$ (see Ref.
\onlinecite{contivignale} for details; the full elastic constants $\mu$ and
$K$ given
there also include the kinetic and the ALDA part of the xc contribution).

Because ${\rm Im} f_{\xc, L(T)}(\omega)$ vanish linearly for $\omega \to 0$
(point 4 of Section II.B), we see that  the viscosity coefficients
$\tilde{\eta}$ and $\tilde{\zeta}$ stay finite in the $\omega \to 0$ limit.
Eqs.  (\ref{imzerolimit1},\ref{imzerolimit2}) of Section II.B imply that the
$\omega \to 0$
limit of the  bulk viscosity $\lim_{\omega \to 0} \zeta (\omega)$ is
exactly zero at least
to within the accuracy of our perturbative calculation. By virtue of the
limiting form
(\ref{zerolimit}), we also see that $\lim_{\omega \to 0} K_{\rm dyn}
(\omega) = 0$,
implying that the bulk modulus of the electron liquid is entirely accounted
for by the
ALDA contribution $\epsilon_{\xc}^{\prime \prime}$.

The fact that the  shear modulus does not vanish for $\omega \to 0$ but
tends to the
finite value  of  Eq. (\ref{zerolimit}) is perhaps surprising.  One
ordinarily thinks
of liquids as having zero shear modulus.  The reason for this strange
behavior is that
we are taking the $k \to 0$ limit {\it before} the $\omega \to 0$ limit.
Thus, the system
remains ``dynamical'' down to zero frequency. Of course, this would not be
true if the
$\omega \to 0$ limit were taken at finite $k$. In that  limit,
$f^h_{\xc,T}$ is no
longer related to the shear modulus, but to the static diamagnetic
susceptibility, which
is extremely small. The truly static shear modulus is zero, as expected.

\subsection {The exchange-correlation field in the inhomogeneous electron gas}
The main result of the previous section, Eq. (\ref{exc5}), is artfully written
so that it can immediately be turned into a local density approximation for
the xc electric field of an inhomogeneous electron liquid through the
replacement
$n \to n_0({\bf r})$,  where $n_0({\bf r})$ is the ground-state density of the
inhomogeneous liquid. Of course, the xc kernels must also be evaluated at
the local
density.

An important question is this: Why should the replacement $n \to n_0({\bf r})$
be done in Eq. (\ref{exc5}) rather than in one of the many equivalent
expressions one can generate starting from Eq. (\ref{exc2})? For example,
why not write the second term on the right-hand side of Eq. (\ref{exc3})
in the equivalent form
\begin{equation} \label{exc3alt}
\frac{1}{i \omega}  \Big\{ \left [f^h_{\xc,L}(\omega)  -
f^h_{\xc,T}(\omega) - \epsilon_{\xc}'' \right ]  \nabla(\nabla
\cdot {\bf j}_1) + f^h_{\xc,T}(\omega) \nabla^2 {\bf j}_1 \Big\}
\end{equation}
{\it before} substituting $n$ by $n_0({\bf r})$? The answer is that this and
similar ambiguities are completely removed by general physical requirements
which we now discuss.

First of all, because  the Coulomb interaction obeys Newton's third law,
the net force exerted by the xc electric field on the system must vanish.
At the local
level, Newton's third law implies that a small volume of the electron
liquid cannot
exert a net force on itself. Accordingly, the net force acting on an
arbitrary volume
element must be expressible as the integral of the {\it external} stresses
exerted by the
surrounding fluid on the surface of the volume element. The mathematical
expression
of this requirement is that the force density must be the divergence of a
local stress
tensor as in Eq. (\ref{exc5}).

A similar argument can be applied to the net {\it torque} acting on a volume
element of the fluid.  Again, this must be expressible in terms of a
surface integral,
and it is not difficult to see that the condition  for this to happen is
that the stress
tensor  be a symmetric rank-2 tensor.\cite{landaulifshitz}

Finally, Galilean invariance requires the stress tensor to vanish
identically when the fluid moves as a whole, i.e., when the velocity field
is  spatially
uniform.  It is  for this reason that the stress tensor must contain
derivatives of
the velocity field and not of the current.

These criteria unambiguously establish Eq.~(\ref{exc5}) as the correct
expression in which the substitution $n \to n_0(\vec r)$ should be made.
This expression was originally derived by Vignale
and Kohn (VK)  by a more laborious and apparently quite different path,
which more
clearly exposed the underlying approximations and the conditions for their
validity. VK considered a weakly inhomogeneous electron
liquid modulated by a charge-density wave of small amplitude
$\gamma$ and small wave vector $\bf q$. Both $k$ and $q$ were
assumed to be small not only relative to the Fermi momentum $k_F$
but also relative to $\omega/u_F$ ($u_F$ being the Fermi
velocity). The latter condition assures that the phase velocity of
the density disturbance is much faster than the Fermi velocity, so
that no form of static screening can occur. Under these
assumptions, all the components of the tensorial kernel
$f_{\xc,\alpha\beta}$ could be calculated, up to first order in
the amplitude of the charge density wave, and to second order in
the wave vectors $k$ and $q$. The calculations were greatly
facilitated by a set of sum rules which are mathematically
equivalent to the zero-force and zero-torque requirements
discussed above. The result of the analysis was that the diagonal
matrix elements $f_{\xc,\alpha\beta}({\bf k},{\bf k},\omega)$
remain equal to $f^h_{\xc}(k,\omega)$ to first order in $\gamma$,
but the off-diagonal elements $f_{\xc,\alpha\beta} ({\bf k}+{\bf
q},{\bf k},\omega)$ acquire a finite value, given by \cite{VK}
\begin{widetext}
\begin{eqnarray} \label{fxcoffdiagonal}
f_{\xc,\alpha\beta}({\bf k} + {\bf q}, {\bf k},\omega)&=&
-\frac{\gamma}{\omega^2} \bigg\{
(\delta f^h_{\xc L} - f^h_{\xc T}) q_\alpha q_\beta +
f^h_{\xc T} q^2 \delta_{\alpha\beta} - n \: \frac{\partial f^h_{\xc T}}
{\partial n} \: {\bf k} \cdot({\bf k}+{\bf q})\delta_{\alpha\beta}
\nonumber\\
&& {}+ {\cal A}(n,\omega)(k_\alpha + q_\alpha)k_\beta
- {\cal B}(n,\omega)k_\alpha(k_\beta + q_\beta) \bigg\},
\end{eqnarray}
\end{widetext}
where $\delta f^h_{\xc L}\equiv f^h_{\xc L}(\omega,n)-\epsilon_{xc}^{\prime
\prime}(n)$, ${\cal A}(n,\omega) \equiv [n(2\partial f^h_{\xc T}
/\partial n - \partial f^h_{\xc L}/
\partial n + 3f^h_{\xc T} - \delta f^h_{\xc L}]$ and ${\cal B}(n,\omega) \equiv
[n\partial f^h_{\xc T}/ \partial n + 3f^h_{\xc T} -\delta f^h_{\xc L}]$.
A remarkable feature of this result is that the off-diagonal matrix elements
of $f_{\xc,\alpha\beta}$ do not exhibit any singularity for $k$ or $q$ tending
to zero in any order. This is in marked contrast with the off-diagonal elements
of the scalar (density) xc kernel which, when calculated for the same system,
exhibit a power singularity of the form ${\bf k} \cdot {\bf q}/k^2$ for
$k \to 0$ at finite $q$. This is the fundamental reason why the dynamical
local density approximation is possible in terms of the current, but not in
terms of the density.

Eq. (\ref{fxcoffdiagonal}) can be translated into a real space expression for
${\bf E}_{\xc,1}({\bf r},\omega)$. More details of the derivation, which is
quite laborious, are given in Ref. \onlinecite{VK}. Finally, the resulting
expression can be rearranged\cite{VUC} in the elegant form of Eq. (\ref{exc5}).
The conditions of validity of the real-space approach are
$|\nabla n_0({\bf r})|/n_0({\bf r})$ much smaller than $ k_F({\bf r})$ and
$\omega/u_F({\bf r})$, where $k_F({\bf r})$ and $u_F({\bf r})$ are the
local Fermi
momentum and velocity. In Ref. \onlinecite{ullrichvignale}, the practical
relevance of
these conditions was investigated in detail. It was found that the
approach could be successfully applied to describe intersubband plasmons
in wide single quantum wells, but failed for narrow double quantum wells.
The failure in the latter case was traced back to a strong violation
of the above criteria of validity in the region of the barrier between the
two wells. The physical reason is that electronic motion through tunnelling
barriers implies strong internal compression of the electron liquid, which
locally destroys coherence of the electron dynamics and leads to a breakdown
of the simple hydrodynamical picture. In such a situation, a hybrid between
the VUC and the more robust Gross-Kohn approximation (\ref{GK})
provides a pragmatic and practically useful remedy. \cite{ullrichvignale}

Because the occurrence of two spatial derivatives of the velocity field in the
post-ALDA term is dictated by general principles, Eq. (\ref{exc5}) is expected
to remain valid even for large values of $u$, i.e., in the nonlinear regime,
provided that $u$ and $n$ are sufficiently slowly varying. The argument goes
as follows. Suppose we tried to extend Eq. (\ref{exc5}) into the nonlinear
regime by including terms of order $u^2$. Because the stress tensor must
depend on first derivatives of $u$, such corrections would have to go as
$(\nabla u)^2$. But then the force density, given by the derivative of the
stress tensor, would have to involve at least three derivatives. Thus, for
sufficiently small spatial variation of the density and velocity fields, the
nonlinear terms can be neglected.

Since the ALDA is an intrinsically nonlinear approximation, VUC proposed that
Eq. (\ref{exc5}), written in the time domain, could provide an appropriate
description of both linear and nonlinear response properties. A nonlinear,
retarded expression for $v_{\xc,1}$ was also proposed by Dobson {\em et al.}
\cite{bunner} The two approximations coincide in ``one-dimensional systems"
(i.e.,
when one has a unidirectional current density field that depends only on one
coordinate), but differ in the general case.

\section{Memory function formalism and TCDFT for homogeneous systems}
\label{secIII}

In the preceding sections, we outlined a linear response formalism
within TCDFT that goes beyond the adiabatic approximation and
allows one to account for intrinsic damping of collective
excitations in electronic systems, caused by dynamical many-body
effects. As mentioned in the Introduction, this is usually not
sufficient to achieve quantitative agreement with experimentally
measured linewidths. In reality, intrinsic damping is often
overshadowed by strong extrinsic dissipation mechanisms, such as
impurities or disorder (in this paper, we consider the
low-temperature case only and limit the discussion to systems
where LO phonon scattering does not occur).

Effects of impurities and disorder in the linear dynamics of a
many-electron system are conveniently discussed in the language of
relaxation functions,\cite{kadanoff} which then naturally leads to the so-called
memory function formalism. \cite{forster,yoshida} In this paper,
we perform a conceptually new step and unite the memory function
formalism with TCDFT in the linear response regime, which will
then allow us to treat both intrinsic and extrinsic damping from
first principles and on an equal footing. This is necessary for an
accurate description of experiments performed on very clean
samples (such as the quantum well we shall discuss in section
\ref{secV}), where intrinsic and extrinsic damping may be of
comparable magnitude.

The purpose of this section is twofold: To make
this paper self-contained, we first review the
memory function formalism for the homogeneous case.
We then make contact between
relaxation and response functions, thereby integrating the memory
function formalism with TCDFT. In Section \ref{secIV} we shall
extend the approach to systems that are inhomogeneous in one
spatial direction (such as quantum wells), and show how it can be
applied to discuss intrinsic and extrinsic damping of collective
charge-density excitations in such systems.

\subsection{Relaxation and Linear Response}
\label{subsecIIIA}

Suppose we are interested in the dynamics of a set of $N+1$ observables,
$\{ \hat{A}_0({\bf r},t), \ldots, \hat{A}_N({\bf r},t)\}$, each
$\hat{A}_i({\bf r},t)$ coupling to a small perturbing external
field $\delta a_i({\bf r},t)$. The Hamiltonian in the presence of
these fields is given (in Schr\"odinger representation) by
\begin{equation} \label{3.A.1}
\hat{H}(t) = \hat{H}_0 + \sum_i \int \! d^3r \: \hat{A}_i({\bf r})
\delta a_i({\bf r},t) \;.
\end{equation}
Consider now the case where the external fields are
adiabatically turned on beginning at $t=-\infty$, and then abruptly switched
off at $t=0$:
\begin{equation} \label{3.A.2}
\delta a_i({\bf r},t) = \left\{
\begin{array}{cc} \delta a_i({\bf r})\:e^{\eta t} & \hspace{0.5cm} \mbox{for}\;t\le 0 \\
0 & \hspace{0.7cm} \mbox{for}\;t>0 \;. \end{array} \right.
\end{equation}
The system starts out from non-equilibrium at $t=0^+$ and, being
left to itself, relaxes back towards equilibrium.  The first-order
change of the non-equilibrium expectation value of an observable
can then be written as \cite{kadanoff,forster,yoshida,gotze}
\begin{equation} \label{3.A.3}
\delta\langle \hat{A}_i({\bf r},t)\rangle_{\rm noneq.} =
 \sum_j \int\!d^3 r' \: \tilde{C}_{ij}({\bf r},{\bf r}',t)\:
\delta a_j({\bf r}') \;,
\end{equation}
where the correlation (or Kubo) function in the presence of disorder
is defined as
\begin{eqnarray} \label{3.A.4}
\tilde{C}_{ij}({\bf r},{\bf r}',t) &=& \left \langle\int_0^{\beta}
\!d\beta' \Big[ \langle \hat{A}_i({\bf r},t) \hat{A}_j({\bf
r'},-i\beta')\rangle_{\rm eq.} \right.
\nonumber\\
&-& \left.\langle \hat{A}_i({\bf r},t)\rangle_{\rm eq.} \langle
\hat{A}_j({\bf r'},-i\beta')\rangle_{\rm eq.} \Big]
\rule[0mm]{0mm}{6mm} \right\rangle_{\rm disorder} .
\end{eqnarray}
This may be rewritten as
\begin{equation} \label{3.A.5}
\tilde{C}_{ij}({\bf r},{\bf r}',t) = \left \langle \hat{A}_i({\bf
r}) \left| e^{-i {\cal L} t} \right| \hat{A}_j({\bf r'}) \right
\rangle\;,
\end{equation}
where $\cal L$ is the Liouville operator governing time evolution of the
system via
\begin{equation}\label{3.A.5a}
\:\dot{ \!\!\hat{A}}_i(t) =i{\cal L} \hat{A}_i(t) =
[\hat{A}_i(t),\hat{H}]/(i \hbar),
\end{equation}
and the scalar product $\langle \ldots | \ldots \rangle$ is
defined by Eq. (\ref{3.A.4}). We impose normalization on the set
of variables, i.e. $\langle \hat{A}_i({\bf r}) | \hat{A}_j({\bf
r'}) \rangle = \delta_{ij}$. In the following, we are interested
in the zero temperature limit ($\beta\to \infty$).

The correlation function (\ref{3.A.4}) is related to the absorptive part of
the usual quantum mechanical response functions $\chi_{ij}({\bf r},{\bf r}',t)$
as follows:
\begin{equation} \label{3.A.6}
i \partial_t \tilde{C}_{ij}({\bf r},{\bf r}',t) =
\chi_{ij}''({\bf r},{\bf r}',t) \;.
\end{equation}
Since relaxation occurs for $t>0$, it is customary to introduce
the Laplace transform of
the quantities of interest, which, e.g., for the correlation function
is defined as
\begin{equation} \label{3.A.7}
\tilde{C}_{ij}({\bf r},{\bf r}', \xi) = \int_0^\infty \!dt \: e^{i\xi t}
\: \tilde{C}_{ij}({\bf r},{\bf r}',t) \;,
\end{equation}
where $\xi$ is a complex number in the upper half of the complex plane.
One then finds the following relationship between the Kubo relaxation functions
and the response functions:
\begin{equation} \label{3.A.8}
\tilde{C}_{ij}({\bf r},{\bf r}', \xi) = \left[ \chi_{ij}({\bf r},{\bf r}', \xi)
- \chi_{ij}({\bf r},{\bf r}', i0)\right]/(i\xi) \;.
\end{equation}

For the remainder of this section, we assume that the system is spatially
homogeneous (a more general case will be considered in Sec. \ref{secIV}).
Eq. (\ref{3.A.3}) can then be Fourier transformed into momentum
space, and one obtains
\begin{equation} \label{3.A.9}
\delta\langle \hat{A}_i({\bf q},\xi)\rangle_{\rm noneq.} = \sum_j
\tilde{C}_{ij}({\bf q},\xi) \: \delta a_j({\bf q}) \;.
\end{equation}
The Laplace transform of the correlation function (\ref{3.A.5}) is then given by
\begin{equation}\label{3.A.10}
\tilde{C}_{ij}({\bf q},\xi) = \left \langle \hat{A}_i({\bf q})
\left| \frac{i}{\xi - {\cal L}} \right| \hat{A}_j({\bf q}) \right
\rangle .
\end{equation}

\subsection{Projectors and Memory Functions}
\label{subsecIIIB} The observables $\{\hat{A}_0({\bf
q}),\ldots,\hat{A}_N({\bf q})\}$ can be regarded as vectors in a
Hilbert space. The Liouvillian $\cal{L}$  acts as a linear
operator in that space, see Eq. (\ref{3.A.5a}). We define a
projection operator $\cal P$ onto the space spanned by
$\{\hat{A}_0({\bf q}),\ldots,\hat{A}_N ({\bf q})\}$ as
\begin{equation}\label{3.B.1}
{\cal P} = \sum_i |\hat{A}_i({\bf q})\rangle \langle
\hat{A}_i({\bf q})| \;,
\end{equation}
and its complement ${\cal Q} \equiv 1- {\cal P}$
projects perpendicular to it. One can then formally write
Eq. (\ref{3.A.10}) as
\begin{equation}\label{3.B.2}
\tilde{C}_{ij}({\bf q},\xi) = \left \langle \hat{A}_i({\bf q})
\left| \frac{i}{\xi - {\cal LQ - LP}} \right| \hat{A}_j({\bf q})
\right \rangle .
\end{equation}
Following Forster, \cite{forster} one performs a few
straightforward manipulations in Eq. (\ref{3.B.2}) and finds
\begin{eqnarray}\label{3.B.3}
\lefteqn{ \tilde{C}_{ij}({\bf q},\xi) = \frac{i}{\xi}\:
\delta_{ij} + \frac{1}{\xi} \sum_k \left\{ \langle \hat{A}_i({\bf
q})| {\cal L} | \hat{A}_k({\bf q}) \rangle
\rule[0mm]{0mm}{5mm}\right.}
\nonumber\\
&-&\!\!\left.\left \langle \:\dot{ \!\!\hat{A}}_i({\bf q}) \left|
{\cal Q}\: \frac{1}{{\cal QLQ}- \xi}\:{\cal Q} \right| \:\dot{
\!\!\hat{A}}_k({\bf q}) \right \rangle \right\}
\tilde{C}_{kj}({\bf q},\xi) .
\end{eqnarray}
Defining
\begin{eqnarray}
C_{ij}({\bf q},\xi) &=& i \tilde{C}_{ij}({\bf q},\xi)\\
\Omega_{ik}({\bf q}) &=& \langle \hat{A}_i({\bf q})| {\cal L} | \hat{A}_k({\bf q})\rangle\\
M_{ik}({\bf q},\xi) &=& \left \langle \:\dot{ \!\!\hat{A}}_i({\bf
q}) \left| {\cal Q}\: \frac{1}{{\cal QLQ} - \xi}\:{\cal Q} \right|
\:\dot{ \!\!\hat{A}}_k({\bf q}) \right \rangle \label{3.B.3a}
\end{eqnarray}
we rewrite Eq. (\ref{3.B.3}) as follows:
\begin{equation}\label{3.B.4}
\sum_k \Big\{ \xi \delta_{ik} - \Omega_{ik}({\bf q})
+ M_{ik}({\bf q},\xi)\Big\}C_{kj}({\bf q},\xi) = -\delta_{ij} \;.
\end{equation}
$\Omega_{ij}({\bf q})$ can be viewed as characteristic frequency
or restoring force matrix of the system. Our particular interest,
however, lies in the memory function matrix $M_{ik}({\bf q},\xi)$,
which introduces dissipation into the electron dynamics of the
system. In general, dissipation originates from {\em intrinsic} as
well as {\em extrinsic} scattering mechanisms. The former, caused
by electron-electron interaction alone, are present even in a
perfectly ``clean'' system.\cite{HPTfootnote} In the previous
section we discussed the treatment of intrinsic dissipation in the
framework of TCDFT. Now we describe how additional extrinsic
dissipation (e.g. caused by scattering off disorder or charged
impurities) can be taken into account simultaneously.

$M_{ik}$ has the formal structure of a correlation function
between two projected forces, ${\cal Q}\:\dot{ \!\!\hat{A}}_i({\bf
q})$ and ${\cal Q} \:\dot{ \!\!\hat{A}}_j({\bf q})$. These forces
act perpendicular to the vector space of variables $\{\hat{A}_i\}$, thus providing
a coupling to other degrees of freedom of the system (which
effectively form a ``thermal bath''). Accordingly, the frequency
dynamics of $M_{ik}({\bf q},\xi)$ is determined by $\cal QLQ$,
where those fluctuations of the Liouville operator are projected
out that occur only within the space of variables $\{\hat{A}_i\}$,
therefore describing the internal dynamics of the ``bath''.

The correlation functions $C_{ij}({\bf q},\xi)$ are determined by
a set of $(N+1)^2$ coupled equations, Eq. (\ref{3.B.4}), whose
solution will be discussed for an example in Sec.
\ref{subsecIIIC}. The observables ${\hat{A}_i}$ are not restricted
to be scalars, but can also be vectors (or $n$th-rank tensors). In
general, all correlation functions $C_{ij}(k,\omega)$ as well as
$\Omega_{ij}$ and $M_{ij}$ are tensors whose rank equals the
sum of the ranks of ${\hat{A}_i}$ and ${\hat{A}_j}$.

Assuming that explicit solutions for the
$C_{ij}$ have been found, the final step is then to make contact
with linear response theory, which involves Fourier transforms (with
frequency $\omega$) rather than Laplace transforms (with frequency $\xi$) of
the associated response and correlation functions. Fortunately,
the relationship between Fourier and Laplace transforms is a straightforward
linear one, so that the response functions $\chi_{ij}({\bf q},\omega)$ can be
simply obtained from
\begin{equation} \label{3.B.5}
C_{ij}({\bf q}, \omega) = \left[ \chi_{ij}({\bf q}, \omega)
- \chi_{ij}({\bf q}, 0)\right]/\omega \;.
\end{equation}
Here, $\omega$ are real frequencies, and
$C_{ij}({\bf q}, \omega) \equiv C_{ij}({\bf q}, \xi = \omega + i 0^+)$.

\subsection{Generalized Relaxation Time Approximation}
\label{subsecIIIC} In the following, we discuss the case where
there are only two observables of interest: density fluctuations
$\hat{\rho}({\bf q})$ and current density $\hat{\bf j}({\bf q})$.
Including normalization, we have
\begin{equation}\label{3.C.1}
\hat{A}_0({\bf q}) = \frac{\hat{\rho}({\bf q})}{\langle
\hat{\rho}({\bf q})|\hat{\rho}({\bf q}) \rangle ^{1/2}}
\end{equation}
and
\begin{equation}\label{3.C.2}
\hat{A}_1({\bf q}) = \frac{ \hat{\bf j}({\bf q})}{\sqrt{n}} \;,
\end{equation}
where $n$ is the uniform density of the system, and $\langle
\hat{\rho}({\bf q})|\hat{\rho}({\bf q})\rangle = \chi({\bf q},0)$,
the static density-density response function (in the presence of
disorder). Eq. (\ref{3.B.4}) then describes the four associated
correlation functions $C_{00}$, $C_{01}$, $C_{10}$ and $C_{11}$,
i.e. $\hat{\rho}$-$\hat{\rho}$, $\hat{\rho}$-$\hat{\bf j}$,
$\hat{\bf j}$-$\hat{\rho}$ and $\hat{\bf j}$-$\hat{\bf j}$.

Since we deal with the homogeneous and isotropic case, Eq.
(\ref{3.B.4}) decouples into two independent equations for the
longitudinal ($L$) and transverse ($T$) components of the
correlation functions:
\begin{equation} \label{3.B.8}
\sum_{k=0}^1\Big[ \omega \delta_{ik} - \Omega_{ik}^L({\bf q}) +
M_{ik}^L({\bf q},\omega) \Big] C_{kj}^L({\bf q},\omega) = -\delta_{ij}\;,
\end{equation}
\begin{equation}\label{3.B.9}
\Big[ \omega   - \Omega_{11}^T({\bf q}) + M_{11}^T({\bf q},\omega)
\Big] C_{11}^T({\bf q},\omega) = -1 \;.
\end{equation}
Eq. (\ref{3.B.8}), with $i,j=0,1$, represents a system of 4
equations coupling the 4 possible longitudinal correlation
functions ($\hat{\rho}$-$\hat{\rho}$, $\hat{\rho}$-$\hat{j}^L$,
$\hat{j}^L$-$\hat{\rho}$ and $\hat{j}^L$-$\hat{j}^L$). Since there
is no coupling between density and transverse currents, there is
only a single transverse correlation function,
$\hat{j}^T$-$\hat{j}^T$, determined by Eq. (\ref{3.B.9}). Using
the continuity equation,
\begin{equation} \label{3.C.3}
{\cal L} \hat{\rho}({\bf q}) = -q \hat{j}^L({\bf q}) \;,
\end{equation}
we convince ourselves that indeed $\langle \hat{A}_0 | \hat{A}_1^L
\rangle = 0$. Furthermore, since ${\cal L} \hat{A}_0$ is
proportional to $\hat{A}_1^L$, the first component of the
``perpendicular'' force ${\cal Q }\:\dot{ \!\!\hat{A}}_0$ is
identically zero, so that
\begin{equation}\label{3.C.4}
M_{00}^L({\bf q},\omega)=M_{01}^L({\bf q},\omega) = M_{10}^L({\bf q},\omega) = 0 \;.
\end{equation}

In the following, we will be concerned with the limit of {\em weak
disorder}. In this limit, it is a good approximation to assume
that all {\em static} correlation functions are unaffected by
disorder. This means that $\Omega^{L(T)}({\bf q})$ contains
effects of Coulomb interaction only. Likewise, we assume
\begin{equation}
\chi_{ij}^{L(T)}({\bf q},0) = \chi_{ij}^{Lc(Tc)}({\bf q},0) \;,
\end{equation}
where the superscript ``$c$'' denotes the ``clean'' response
function. In general, static disorder effects (mainly changes of
the density of state at the Fermi surface) are weak, provided
$(\tau \varepsilon_F)^{-1}\ll1$, where $\varepsilon_F$ is the
Fermi energy, and $\tau$ is a characteristic disorder scattering
time for the system under study.

In general, the memory functions contain both intrinsic (Coulomb
interaction) and extrinsic (disorder) contributions. In the limit
of weak disorder, one can separate them as follows:
\begin{equation}
M^{L(T)}_{11}({\bf q},\omega) \equiv M^{L(T)}_{\rm in}({\bf
q},\omega) + M^{L(T)}_{\rm ex}({\bf q},\omega).
\end{equation}
We combine the intrinsic part with $\Omega_{ik}$, thereby defining
a frequency-dependent, dissipative restoring force matrix
$\Omega^{{\rm in},L(T)}_{ik}({\bf q},\omega)$, which contains
effects of Coulomb interactions only:
\begin{equation}
\Omega^{{\rm in},L(T)}_{ik}({\bf q},\omega) \equiv
\Omega^{L(T)}_{ik}({\bf q}) + M^{L(T)}_{\rm in}({\bf q},\omega)
\delta_{i1} \delta_{k1}.
\end{equation}
One then obtains from Eqs. (\ref{3.B.8}) and (\ref{3.B.9}):
\begin{widetext}
\begin{equation} \label{3.C.7}
\sum_{k=0}^1\Big\{ [\omega+M^L_{\rm ex}({\bf
q},\omega)]\:\delta_{ik} - \Omega_{ik}^{{\rm in},L}({\bf
q},\omega) \Big\} C_{kj}^L({\bf q},\omega) = -\delta_{ij} +
\delta_{i0} M^L_{\rm ex}({\bf q},\omega) C_{ij}^L({\bf q},\omega)
\;, \quad i,j =0,1
\end{equation}
\end{widetext}
\begin{equation}\label{3.C.8}
\Big[ \omega+M^T_{\rm ex}({\bf q},\omega) - \Omega_{11}^{{\rm
in},T} ({\bf q},\omega) \Big] C_{11}^T({\bf q},\omega) = -1 \;.
\end{equation}
For notational brevity, we suppress the $({\bf q},\omega)$
dependence and the subscript ``ex'' of $M^L_{\rm ex}$ and $M^T_{\rm ex}$ in the
following. To solve Eqs. (\ref{3.C.7}) and (\ref{3.C.8}), we
introduce ``clean'' longitudinal (transverse) reference functions
$C_{ij}^{Lc(Tc)}({\bf q}, \omega)$, defined in the absence of
dissipation (i.e., $M^L = 0 = M^T$), as follows:
\begin{equation} \label{3.C.9}
\sum_{k=0}^1\Big[ \omega\:\delta_{ik} - \Omega_{ik}^{{\rm
in},L}({\bf q},\omega) \Big] C_{kj}^{Lc}({\bf q},\omega) = -\delta_{ij}
\;, \quad i,j = 0,1\;,
\end{equation}
\begin{equation}\label{3.C.10}
\Big[ \omega - \Omega_{11}^{{\rm in},T}({\bf q},\omega) \Big]
C_{11}^{Tc}({\bf q},\omega) = -1 \;.
\end{equation}
The desired correlation functions are then expressed in terms of these
reference functions.
In the longitudinal case, we find
\begin{eqnarray} \label{3.C.11}
C_{ij}^L({\bf q},\omega)&=& C_{ij}^{Lc}({\bf q},\omega+M^L) -
C_{i0}^{Lc}({\bf q},\omega+M^L) \nonumber\\
&& \times M^L \, C_{0j}^L({\bf q},\omega) \;, \quad i,j = 0,1\;,
\end{eqnarray}
or explicitly
\begin{eqnarray}
C_{00}^L({\bf q},\omega) &=& \frac{C_{00}^{Lc}}{1+M^L
C_{00}^{Lc}} \label{3.C.13}\\
C_{10}^L({\bf q},\omega) &=& \frac{C_{10}^{Lc}}{1+M^L
C_{00}^{Lc}} \label{3.C.14}\\
C_{01}^L({\bf q},\omega) &=& \frac{C_{01}^{Lc}}{1+M^L
C_{00}^{Lc}} \label{3.C.15}\\
C_{11}^L({\bf q},\omega) &=& \frac{C_{11}^{Lc} + C_{11}^{Lc} M^L
C_{00}^{Lc} - C_{10}^{Lc} M^L C_{01}^{Lc} }{1+M^L C_{00}^{Lc}}
\label{3.C.16} ,
\end{eqnarray}
where all ``clean'' functions carry the arguments $({\bf q},\omega+M^L)$.
For the transverse case, on the other hand,  we simply obtain
\begin{equation} \label{3.C.12}
C_{11}^T({\bf q},\omega) = C_{11}^{Tc}({\bf q},\omega+M^T) \;.
\end{equation}

We now make contact with the response functions as follows. For the
case of density-density response, we have from Eq. (\ref{3.B.5}):
\begin{equation} \label{3.D.1}
C_{00}^L({\bf q},\omega) = \frac{1}{\omega} \Big[ \chi_{00}^L({\bf q},\omega)
-  \chi_{00}^L({\bf q},0)\Big] \Big/\chi_{00}^L({\bf q},0)\;,
\end{equation}
where the function $\chi_{00}^L({\bf q},0)$ in the denominator arises from the normalization
of the variable $\hat{A}_0$, see Eq. (\ref{3.C.1}). Similar expressions are obtained
for the other longitudinal and transverse correlation functions.
  We then find from Eq. (\ref{3.C.13}):
\begin{eqnarray} \label{3.D.2}
\frac{1}{\chi_{00}^L({\bf q},\omega)}&=& \frac{\omega}{(\omega +
M^L)} \frac{1}{\chi_{00}^{Lc}({\bf q},\omega+M^L)} \nonumber\\
&+& \frac{M^L}{(\omega + M^L)}\frac{1}{\chi_{00}^{Lc}({\bf q},0)}
\;.
\end{eqnarray}
Eq. (\ref{3.D.2}) is formally in agreement with Belitz and Das Sarma
[equation (2.3) in Ref. \onlinecite{belitz}].
However, $\chi^c_{00}(\qqp,\omega)$ here
denotes the {\em exact}, fully interacting response function of the
homogeneous electron gas, not just the RPA response function.

>From (\ref{3.D.2}), one easily derives explicit expressions for
longitudinal density-current and current-current response functions
using
\begin{equation} \label{3.D.3}
\chi_{00}^L({\bf q},\omega) = \frac{q}{\omega}\chi^L_{10}({\bf
q},\omega) = \frac{q}{\omega}\chi^L_{01}({\bf q},\omega) =
\frac{q^2}{\omega^2}\chi^L_{11}({\bf q},\omega) .
\end{equation}
It is not difficult to show that the results for $\chi^L_{10}$,
$\chi^L_{01}$ and $\chi^L_{11}$ obtained in this fashion are consistent with
Eqs.  (\ref{3.C.14})--(\ref{3.C.16}).

In the same way one finds the transverse current-current response function
from Eq. (\ref{3.C.12}):
\begin{eqnarray}\label{3.D.4}
\chi_{11}^T({\bf q},\omega) &=& \frac{\omega}{(\omega + M^T)} \:
\chi_{11}^{Tc}({\bf q},\omega + M^T)\nonumber\\
& +& \frac{M^T}{(\omega + M^T)} \: \chi_{11}^{Tc}({\bf q},0) \;.
\end{eqnarray}
Expressions that are formally similar to Eqs.
(\ref{3.D.2})-(\ref{3.D.4}) were recently derived by Conti and
Vignale\cite{contivignale}  in the framework of Mermin's
relaxation time approximation. \cite{mermin} In this formalism, the
role of $M({\bf q},\omega)$ is taken by a frequency- and
momentum-independent phenomenological scattering rate $i/\tau$.
Note that the second term on the right-hand side of (\ref{3.D.4})
is absent in Ref. \onlinecite{contivignale}, because there the
diamagnetic susceptibility of the electron gas,
$\chi_{11}^{Tc}({\bf q},0)$, was implicitly taken to be zero.

Finally, explicit expressions for the memory functions $M^L({\bf
q},\omega)$ and $M^T({\bf q},\omega)$ are obtained in the
following way from Eq. (\ref{3.B.3a}): first, we approximately
write
\begin{equation}
\label{3.D.5} M^L({\bf q},\omega) = \left\langle \hat{\bf
F}^L({\bf q}) \left| \frac{1}{{\cal L} - \omega} \right| \hat{\bf
F}^L({\bf q}) \right\rangle \;,
\end{equation}
and similar for $M^T$, i.e. we resort to the standard
approximation of replacing the projected by the full Liouville
operator by setting ${\cal Q} \approx 1$ in the denominator. We
thus assume that the huge amount of degrees of freedom in the
thermal bath and their extremely complex time evolution are completely
dominating over the small subspace of observables $\hat{A}_0,\hat{A}_1$
and its relatively well-controlled time evolution.

In turn, the fluctuating longitudinal forces have the form
\begin{equation}
\label{3.D.6} \hat{\bf F}^L({\bf q}) = \frac{1}{\sqrt{n}}
\sum_{\bf k} \frac{{\bf q}[{\bf q}\cdot({\bf k}-{\bf q})]}{q^2}\:
U({\bf q}-{\bf k}) \hat{\rho}({\bf k})\;,
\end{equation}
and the transverse forces are
\begin{equation}
\label{3.D.7} \hat{\bf F}^T({\bf q}) = -\frac{1}{\sqrt{n}}
\sum_{\bf k} \frac{{\bf q}\times({\bf q}\times{\bf k})}{q^2}\:
U({\bf q}-{\bf k}) \hat{\rho}({\bf k})\;.
\end{equation}
Here, $U({\bf q})$ is a random scattering potential. In the
weak-disorder limit, we can perform the following decoupling, up
to within corrections of higher than second order in the disorder
potential:
\begin{widetext}
\begin{equation} \label{3.D.8}
\left\langle U({\bf q}-{\bf k}) \hat{\rho}({\bf k}) \left|
\frac{1}{{\cal L} - \omega} \right|U({\bf q}-{\bf k}') \rho({\bf
k}') \right\rangle \approx \left\langle \hat{\rho}({\bf k}) \left|
\frac{1}{{\cal L} - \omega} \right| \rho({\bf k}') \right\rangle
\Big\langle U({\bf q}-{\bf k}) U({\bf q}-{\bf k}')
\Big\rangle_{\rm disorder}.
\end{equation}
Since the system is homogeneous, we have ${\bf k} = {\bf k}'$,
and we arrive at the following expression for the longitudinal memory
function:
\begin{equation}\label{3.D.9}
M^L({\bf q},\omega) = \frac{1}{n} \sum_{\bf k} \langle U({\bf q}-{\bf k})
\rangle ^2 \:\frac{[{\bf q}\cdot({\bf q}-{\bf k})]^2}{q^2}
\: C^L_{00}({\bf k},\omega)\chi^{Lc}_{00}({\bf k},0) \;.
\end{equation}
One thus needs to calculate $M^L({\bf q},\omega)$ and $C^L_{00}({\bf q},\omega)$
via self-consistent solution of Eqs. (\ref{3.C.13}) and (\ref{3.D.9}).
The so determined $C^L_{00}({\bf q},\omega)$ then serves as input for
the transverse memory function
\begin{equation}\label{3.D.10}
M^T({\bf q},\omega) = \frac{1}{n} \sum_{\bf k} \langle U({\bf q}-{\bf k})
\rangle ^2 \:\frac{[{\bf q}\times({\bf q}\times{\bf k})]^2}{q^4}
\: C^L_{00}({\bf k},\omega)\chi^{Lc}_{00}({\bf k},0) \;,
\end{equation}
which was obtained using the same decoupling approximation that led to
Eq. (\ref{3.D.9}) for $M^L({\bf q},\omega)$.

\section{Memory function formalism and TCDFT for inhomogeneous systems}
\label{secIV}
\subsection{Formalism}
\label{subsecIVA}
We now generalize the memory function formalism to systems that are
inhomogeneous in one spatial direction, but still homogeneous in the plane
perpendicular to it. The example we have in mind are quantum wells whose
direction of growth is the $z$-axis. One can then in general no longer decouple
longitudinal and transverse components of the correlation functions.
A special case where this is still possible will be discussed in some detail
later on.

The generalization of Eq. (\ref{3.B.4}) for this inhomogeneous situation is
\begin{equation}\label{4.A.1}
\sum_k \int\!dz''\:\Big[ \omega \: \delta(z''-z) \: \delta_{ik}
- \Omega_{ik}(\qqp,z,z'') + M_{ik}(\qqp,z,z'',\omega) \Big]
C_{kj}(\qqp,z'',z',\omega) = -\delta_{ij} \: \delta(z-z'),
\end{equation}
where $\qqp$ is the in-plane wave vector.
We again consider density fluctuations and current density as the only
variables. Just like in the homogeneous case, by virtue of the continuity
equation, all elements of the $2\times 2$ memory function matrix $M_{ik}$ are
zero except $M_{11}$. As before, we can separate intrinsic and extrinsic
contributions to the memory function in the weak-disorder limit, 
$M_{11}\equiv M_{\rm in}+M_{\rm ex}$, 
and we combine the intrinsic part with $\Omega_{ik}$, defining 
\begin{equation}
\Omega^{\rm in}_{ik}(\qqp,z,z',\omega) \equiv \Omega_{ik}(\qqp,z,z')+
M_{\rm in}(\qqp,z,z',\omega) \delta_{i1}\delta_{k1}\;.
\end{equation}
Eq. (\ref{4.A.1}) thus becomes
\begin{eqnarray}
\lefteqn{ \hspace{-1.0cm}
\sum_{k=0}^1 \int\!dz''\:\bigg\{ \Big[ \omega \: \delta(z''-z)
+ M_{\rm ex}(\qqp,z,z'',\omega) \Big] \delta_{ik}
- \Omega^{\rm in}_{ik}(\qqp,z,z'',\omega) \bigg\}
C_{kj}(\qqp,z'',z',\omega)} \nonumber\\[3mm]
&=& -\delta_{ij} \: \delta(z-z')
\:+\: \delta_{i0} \int\!dz''\:
M_{\rm ex}(\qqp,z,z'',\omega) C_{ij}(\qqp,z'',z',\omega) \;, \quad i,j=0,1 \;.
\label{4.A.3}
\end{eqnarray}
Similar to Sec. \ref{subsecIIIC}, we will solve this set of equations by
introducing suitable reference functions. However, the inhomogeneity of 
the system prevents us from using the same trick as for the homogeneous
case, where we directly expressed the correlation functions in the presence of disorder
in terms of the ``clean'' correlation functions, with their frequency
argument $\omega$ replaced by $\omega +M_{\rm ex}$. Now, by contrast,
the memory function $M_{\rm ex}(\qqp,z,z',\omega)$
is no longer simply a number, but acts in conjunction with an integral operator, see Eq. (\ref{4.A.3}).
To deal with this difficulty,
we first define a set of intermediate reference functions
$C^{R}_{ij}(\qqp,z,z',\omega)$ that satisfy the following coupled equations:
\begin{eqnarray} \label{4.A.4}
\lefteqn{ \hspace{-4.5cm}
\sum_{k=0}^1 \int\!dz''\:\bigg\{ \Big[ \omega \: \delta(z''-z)
+ M_{\rm ex}(\qqp,z,z'',\omega) \Big] \delta_{ik}
- \Omega^{\rm in}_{ik}(\qqp,z,z'',\omega)
\bigg\} C^{R}_{kj}(\qqp,z'',z',\omega) } \nonumber\\
&&= -\delta_{ij} \: \delta(z-z')\;, \quad i,j=0,1 \;.
\end{eqnarray}
In terms of these reference functions, the full correlation
functions are given, combining Eqs. (\ref{4.A.3}) and (\ref{4.A.4}),
 through the following Dyson-type integral equation:
\begin{eqnarray}
\lefteqn{ C_{ij}(\qqp,z,z',\omega) = C^R_{ij}(\qqp,z,z',\omega) }\nonumber\\
&&{}- \int\!dz_1\! \int \! dz_2 \:
C^R_{i0}(\qqp,z,z_1,\omega) \: M_{\rm ex}(\qqp,z_1,z_2,\omega)
\: C_{0j}(\qqp,z_2,z',\omega) \;.
\label{4.A.5}
\end{eqnarray}
We now use a similar trick to obtain the reference functions $C^R_{ij}$.
Defining the ``clean'' response function via
\begin{equation}\label{4.A.6}
\sum_{k=0}^1 \int\!dz''\:\bigg\{ \omega \: \delta(z''-z) \: \delta_{ik}
- \Omega^{\rm in}_{ik}(\qqp,z,z'',\omega)\bigg\} C^{c}_{kj}(\qqp,z'',z',\omega)
= -\delta_{ij} \: \delta(z-z')\;,
\end{equation}
we get from Eqs. (\ref{4.A.5}) and (\ref{4.A.6}):
\begin{eqnarray}
\lefteqn{ C^R_{ij}(\qqp,z,z',\omega) = C^c_{ij}(\qqp,z,z',\omega)}\nonumber\\
&&{}+ \sum_{k=0}^1 \int\!dz_1\! \int \! dz_2 \:
C^c_{ik}(\qqp,z,z_1,\omega) \: M_{\rm ex}(\qqp,z_1,z_2,\omega)
\: C^R_{kj}(\qqp,z_2,z',\omega) \;.
\label{4.A.7}
\end{eqnarray}
The density-density reference correlation function is explicitly given by
\begin{eqnarray}
\lefteqn{ C^R_{00}(\qqp,z,z',\omega) = C^c_{00}(\qqp,z,z',\omega)}
\nonumber\\
&&{}+ \int\!dz_1\! \int \! dz_2 \:
C^c_{00}(\qqp,z,z_1,\omega) \: M_{\rm ex}(\qqp,z_1,z_2,\omega)
\: C^R_{00}(\qqp,z_2,z',\omega) \nonumber\\
&&{}+ \int\!dz_1\! \int \! dz_2 \:
C^c_{01}(\qqp,z,z_1,\omega) \: M_{\rm ex}(\qqp,z_1,z_2,\omega)
\: C^R_{10}(\qqp,z_2,z',\omega) \;.
\label{4.A.8}
\end{eqnarray}
At this point, it is convenient to introduce the following set of auxiliary
functions, which will allow us later to write the memory function
$M_{\rm ex}$ in a more compact form (see below):
\begin{eqnarray}
\Phi(\qqp,z,z',\omega) &\equiv& \int\! dz''\: C_{00}(\qqp,z,z'',\omega)
\chi^c(\qqp,z'',z',0)\\
\Phi^R(\qqp,z,z',\omega) &\equiv& \int\! dz''\: C^R_{00}(\qqp,z,z'',\omega)
\chi^c(\qqp,z'',z',0)\\
\Phi^c(\qqp,z,z',\omega) &\equiv& \int\! dz''\: C^c_{00}(\qqp,z,z'',\omega)
\chi^c(\qqp,z'',z',0) \;.
\end{eqnarray}
Again it is assumed that all extrinsic damping effects can be neglected for the
static response function.
In terms of these functions, Eq. (\ref{4.A.5}) becomes for $i=j=0$:
\begin{eqnarray} \label{4.A.12}
\lefteqn{
\Phi(\qqp,z,z',\omega) \;=\; \Phi^R(\qqp,z,z',\omega)} \\
&&- \int\!dz_1\!\int\!dz_2\! \int\! dz_3\:
\Phi^R(\qqp,z,z_1,\omega)\: \Big[\chi^c(\qqp,z_1,z_2,0)\Big]^{-1}
M_{\rm ex}(\qqp,z_2,z_3,\omega)\: \Phi(\qqp,z_3,z',\omega) \;.
\nonumber
\end{eqnarray}
In the example to be discussed in section \ref{secV}, only the
case  $\qqp = 0$ will be of interest. Since in this case the continuity
equation can be used to explicitly eliminate the current density in favor of the density,
Eq. (\ref{4.A.8}) can be written as
\begin{eqnarray} \label{4.A.13}
\lefteqn{
\Phi^R(0,z,z',\omega) \;=\; \Phi^c(0,z,z',\omega)} \nonumber\\
&+& \int\!dz_1\!\int\!dz_2\! \int\! dz_3\:
\Phi^c(0,z,z_1,\omega)\: \Big[\chi^c(0,z_1,z_2,0)\Big]^{-1}
M_{\rm ex}(0,z_2,z_3,\omega)\: \Phi^R(0,z_3,z',\omega)\nonumber \\
&+&
\int\!dz_1\! \int \! dz_2
\int_{-\infty}^{z_1} \!d\tilde{z}_1 \:
\chi^c(0,z,\tilde{z}_1,\omega) \:
\frac{M_{\rm ex}(0,z_1,z_2,\omega)}{\sqrt{n(z_1)n(z_2)}}
\int_{-\infty}^{z_2} \!d\tilde{z}_2\:\left[ \omega
\Phi^R(0,\tilde{z}_2,z',\omega) - \chi(0,\tilde{z}_2,z',0)\right]\;.
\end{eqnarray}
The desired density-density response functions are
then finally obtained using
\begin{eqnarray}
\label{4.A.14}
\Phi(\qqp,z,z',\omega) &=& [\chi(\qqp,z,z',\omega) - \chi^c(\qqp,z,z',0)]
/\omega\\
\Phi^R(\qqp,z,z',\omega) &=& [\chi^R(\qqp,z,z',\omega) - \chi^c(\qqp,z,z',0)]
/\omega\\
\Phi^c(\qqp,z,z',\omega) &=& [\chi^c(\qqp,z,z',\omega) - \chi^c(\qqp,z,z',0)]
/\omega \;.
\label{4.A.15}
\end{eqnarray}
To summarize: Eqs. (\ref{4.A.8})--(\ref{4.A.15}) allow one to express the
interacting density-density response function of the system in the presence
of intrinsic {\em and} extrinsic dissipation,
$\chi(\qqp,z,z',\omega)$, in terms of the interacting response function for the
``clean'' system, $\chi^c(\qqp,z,z',\omega)$, i.e. including intrinsic
dissipation alone. $\chi^c$ is calculated, in principle exactly, using the
framework of TCDFT outlined in Sec. \ref{secII}. Although admittedly somewhat
frightening in appearance, the integral equations (\ref{4.A.12}) and
(\ref{4.A.13}) involve only one-dimensional integrals, and their numerical
solution is therefore quite manageable, as will be shown below.

\subsection{Memory functions for impurity and interface roughness scattering} \label{subsecIVB} 

In general, a vector field ${\bf
V}({\bf r})$ is decomposed into longitudinal and transverse
components as follows: ${\bf V}^L({\bf r}) = -(1/4\pi) \nabla \int
d^3 r' [\nabla' \cdot {\bf V}({\bf r}')]/|{\bf r} - {\bf r}'|$ and
${\bf V}^T({\bf r}) = (1/4\pi) \nabla \times \nabla \times \int
d^3r'  {\bf V}({\bf r}') /|{\bf r} - {\bf r}'|$. In our case, i.e.
working in a  mixed $(\qqp,z)$-representation, it is convenient to
define the operator ${\bf D}(\qqp,z)$ as 
\begin{equation} {\bf
D}(\qqp,z)\equiv \left(\begin{array}{c} \qqp\\ i\nabla_z
\end{array} \right) \;.
\end{equation}
The longitudinal fluctuating forces are then given by
\begin{eqnarray} 
\hat{\bf F}^L(\qqp,z) &=& \frac{1}{\sqrt{n(z)}}
\:{\bf D}(\qqp,z)\int\!dz'\!\int \frac{d^2\pp} {(2\pi)^2}
 \: \frac{e^{-\qp|z-z'|}}{2\qp} \nonumber\\
&& \Big\{ {\bf D}(\qqp,z')\cdot \Big[\hat{\rho}(\ppp,z') {\bf
D}(\ppp-\qqp,z') U(\qqp-\ppp,z') \Big]\Big\} \label{4.B.1}
\end{eqnarray}
and the transverse fluctuating forces are
\begin{eqnarray}
\hat{\bf F}^T(\qqp,z) &=& -\frac{1}{\sqrt{n(z)}} \:{\bf
D}(\qqp,z)\times{\bf D}(\qqp,z)\times \int\!dz'\!\int
\frac{d^2\pp}{(2\pi)^2}\: \frac{e^{-\qp|z-z'|}}{2\qp}
 \nonumber\\
&& \hspace{3.0cm} \Big[\hat{\rho}(\ppp,z'){\bf
D}(\ppp-\qqp,z')U(\qqp-\ppp,z')\Big] \;. \label{4.B.2}
\end{eqnarray}
One finds from Eqs. (\ref{4.B.1}) and (\ref{4.B.2})
that in the {\em homogeneous} limit (no $z$ dependence
of the density fluctuations $\rho$ and the scattering potential $U$),
the longitudinal and transverse forces only have in-plane components,
given by the 2-D versions of Eqs. (\ref{3.D.6}) and (\ref{3.D.7}).
In the following, we shall limit the discussion of the {\em inhomogeneous }
situation to a case of special interest, namely $\qqp=0$.
In that case, only the $z$-component of ${\bf F}^L$ survives, and is given by
\begin{equation}
\label{4.B.3} \hat{F}^L_z(z) = \frac{i}{\sqrt{n(z)}} \int
\frac{d^2\pp}{(2\pi)^2} \: \hat{\rho}(\ppp,z) \nabla_{z}
U(-\ppp,z) \;.
\end{equation}
The transverse force, in the same limit, acts in the $x-y$ plane only:
\begin{equation}\label{4.B.4}
\hat{\bf F}^T_{||}(z) = \frac{1}{\sqrt{n(z)}} \int
\frac{d^2\pp}{(2\pi)^2} \: \ppp \: \hat{\rho}(\ppp,z) U(-\ppp,z)
\;.
\end{equation}
We note that, by symmetry, in the limit $\qqp=0$ there is a natural
decoupling of the formalism outlined above (Sec. \ref{subsecIVA})
into separate sets of equations of the type (\ref{4.A.12})--(\ref{4.A.15})
 determining longitudinal and transverse response functions, respectively. In other words,
$L-T$ cross correlations are absent since the associated
fluctuating forces are perpendicular to each other. The longitudinal and transverse 
memory functions are obtained in a quite straightforward manner from 
Eqs. (\ref{4.B.1}) and (\ref{4.B.2}), using the same approximate decoupling procedure
that was used for the homogeneous case in Sec. \ref{subsecIIIC}. The result is:
\begin{equation}\label{4.B.5}
M^L_{\rm ex}(z,z',\omega)= \int \frac{d^2\pp}{(2\pi)^2} \:
\frac{\Phi(\ppp,z,z',\omega)}{\sqrt{n(z) n(z')}}
\:\nabla_z \nabla_{z'} \Big\langle U(-\ppp,z)U(-\ppp,z')\Big\rangle
\end{equation}
and
\begin{equation}\label{4.B.6}
M^T_{\rm ex}(z,z',\omega)= \int \frac{d^2\pp}{(2\pi)^2} \:
\frac{\Phi(\ppp,z,z',\omega)}{\sqrt{n(z) n(z')}}
\:\pp^2\: \Big\langle U(-\ppp,z)U(-\ppp,z')\Big\rangle \;.
\end{equation}
\end{widetext}

The next step consists in finding explicit forms for the disorder-averaged random
scattering potential $U(\ppp,z)$, associated with some extrinsic
damping mechanism. In the following, we shall focus on two examples
specific to quantum wells: damping by charged impurities and by
interface roughness.

The potential associated with a single, statically scree\-ned,
positively charged impurity at position $z_1$ is
\begin{equation}
U(\ppp,z) = \frac{2 \pi }{ \varepsilon(\pp)}
\: \frac{e^{-\pp |z - z_1|}}{\pp} \;,
\end{equation}
where $\varepsilon(\pp)$ is the 2D dielectric function.\cite{ando}
The longitudinal memory function for charged-impurity scattering is thus
\begin{eqnarray} \label{memoryfunction_I}
\lefteqn{
M_{\rm I}^L(z,z',\omega) =
\int \frac{d^2\pp}{\varepsilon^2(\pp)} \:
\frac{\Phi(\ppp,z,z',\omega)}{\sqrt{n(z)n(z')}}
\int\! d\tilde{z} \: n_i(\tilde{z})
}\nonumber\\
&& \times \mbox{sign}(z-\tilde{z})\mbox{sign}(z'-\tilde{z})
e^{-\pp|z-\tilde{z}|} e^{-\pp|z'-\tilde{z}|} ,
\end{eqnarray}
where $n_i(z)$ is the number of impurities per volume.

Likewise, the longitudinal memory function associated with interface
roughness is
\begin{eqnarray}\label{memoryfunction_R}
\lefteqn{
M_{\rm R}^L(z,z',\omega) =
\int \frac{d^2\pp}{(2\pi)^2} \:
\frac{\Phi(\ppp,z,z',\omega)}{\sqrt{n(z)n(z')}}\:
\big\langle U(\pp)^2 \big\rangle}  \nonumber\\
&& \hspace{-6mm} \times \nabla_{\!z} \nabla_{\!z'}\big[
\delta(z-z_l)\delta(z'-z_l)
 + \delta(z-z_r)\delta(z'-z_r) \big] ,
\end{eqnarray}
where $U(\pp)$ is the random roughness scattering potential, assumed for
simplicity to be the same at the left and right interfaces, $z_l$ and $z_r$.
It is common to assume a Gaussian form for the autocorrelation function of
the random interface roughness,\cite{ando,fishman} which leads to
\begin{equation}\label{roughnesspotential}
\big\langle U(\pp)^2 \big\rangle = \pi \mu^2 \Delta^2 \eta^2
e^{-\pp^2 \eta^2/4} \;.
\end{equation}
Here, $\mu$ is the height of the potential step at the interface,
and the correlation length $\eta$ and average roughness height $\Delta$ are
controlled by material and growth conditions.
In the presence of both impurity and roughness scattering, the
memory functions $M_{\rm I}$ and $M_{\rm R}$ are additive (i.e., different
extrinsic scattering mechanisms are assumed to be uncorrelated).

Some practical complications arise from the fact that the memory functions
explicitly depend on $\Phi(\qqp,z,z',\omega)$ at {\em all} $\qqp$, not just $\qqp=0$.
This means that $\Phi$ should be calculated self-consistently from
Eqs. (\ref{4.A.12}) for {\em all} $\qqp$, which is a very
demanding computational task. Therefore, as a first approximation, we
ignore self-consistency and instead use the {\em noninteracting}
$\Phi_0(\qqp,z,z',\omega)$ in (\ref{memoryfunction_I}) and
(\ref{memoryfunction_R}), defined by replacing $\chi$ and $\chi_c$ with
$\chi\ks$ in Eq. (\ref{4.A.14}).
The wave vector dependence of $\Phi_0$ is thus known analytically (see the
explicit expression for $\chi\ks$ below). This is expected to be a
reasonable approximation as long as plasmon damping is not too strong.

\section{Linewidth of intersubband plasmons in a quantum well}
\label{secV}
In semiconductor quantum wells, the
conduction band splits up into several subbands, and electrons
(supplied e.g. by remote doping) can perform collective transiti\-ons between
them. These so-called intersubband (ISB) plasmons
are currently of great experimental and theoretical interest,\cite{books}
being the basis of a variety of new devices operating in the
terahertz regime, such as detectors \cite{detector} and quantum cascade
lasers.\cite{laser} In designing these devices, the emphasis usually lies in
covering a particular frequency range.
However, often it is desirable that the transitions also have a narrow
linewidth, to achieve better frequency resolution and
larger peak absorption in detectors, and higher gain in lasers.
The linewidth arises from a complicated interplay of a variety of
scattering mechanisms, intrinsic (electron-electron and electron-phonon)
as well as extrinsic ones (impurity, alloy-disorder and interface
roughness). Many aspects of this interplay are still not well understood,
in particular the relative importance of the individual mechanisms.\cite{helm}

To disentangle the various contributions
to the ISB linewidth, it is helpful to  consider a
situation where some of them are not effective.
In a recent experiment, Williams {\em et al.} \cite{jon}
studied collective ISB transitions in an n-type
40 nm wide single GaAs/Al$_{0.3}$Ga$_{0.7}$As quantum well, with Si doping
centers 100 nm away from the well. Sharp transitions were found
well below the LO phonon frequency of GaAs ($35.6$ meV), at a temperature of 2.3 K.
Thus, neither remote impurity nor phonon scattering are playing
any significant role (nor is alloy-disorder scattering, as shown in Ref.
\onlinecite{campman}). The linewidth is therefore expected to be dominated by
bulk impurity and interface roughness scattering, while electronic many-body
effects have traditionally been neglected. However, for high-quality samples
such as the one used in the experiment discussed here, this is no longer justified.

In the experiment,\cite{jon} two parameters were controlled independently:
the electronic sheet density $N_s$
(from 0.05 to $1.3 \times 10^{11} \: {\rm cm}^{-2}$),
and a static electric field $E$ perpendi\-cu\-lar to the well
which pushes the electrons against one of its edges. This
provides an ideal tool to distinguish interface roughness
from other damping effects.

\begin{figure}
\unitlength1cm
\begin{picture}(5.0,6.0)
\put(-6.0,-8.8){\makebox(5.0,6.0){
\includegraphics{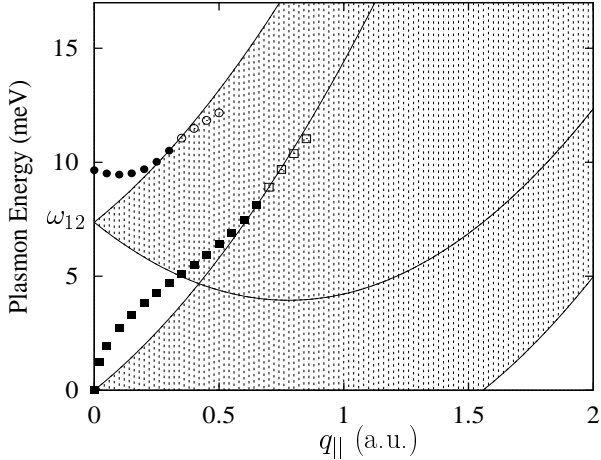}
}}
\end{picture}
\caption{ Dispersions of ISB (circles) and 2D plasmon
(squares) frequencies $\Omega(\qp)$ in a clean quantum well.
$\omega_{12}$ is the difference between the two lowest bare
subband levels. Inside the shaded regions, the plasmons are
subject to strong Landau damping and rapidly die off (open
symbols). The experiment by Williams {\em et al.} \cite{jon} measures the
ISB plasmon frequency and linewidth at $\qp=0$.} 
\label{figure1}
\end{figure}

We describe ISB plasmons within a one-band effec\-tive-mass
approximation with parabolic subbands, \cite{kuznetsov} which is a widely used and, for
our purposes, sufficiently accurate method for
GaAs/Al$_{0.3}$Ga$_{0.7}$As quantum wells. The case of the {\em
clean} quantum well was treated in detail in Ref.
\onlinecite{ullrichvignale}. The non-interacting response function
reads
\begin{eqnarray} \label{chi_ks}
\chi\ks(\qqp,z,z',\omega) &=&
\sum_{\mu=1}^{N_{\rm occ}} \sum_{\nu=1}^{\infty} F_{\mu\nu}(\qp,\omega)
\:\varphi_\mu(z) \varphi_\mu(z') \nonumber\\
&& \times \varphi_\nu (z) \varphi_\nu(z') \:,
\end{eqnarray}
where
\begin{eqnarray} \label{fmunu}
F_{\mu\nu}(\qp,\omega) &=& -2 \int\!  \frac{d^2\kp}{(2\pi)^2}
\left\{
\frac{f(\epsilon_\mu + \kp^2/2)}
{\qqp\kkp + a_{\mu\nu}(\qp) +\omega+i\eta} \right. \nonumber\\
&& \left. {}
+ \frac{f(\epsilon_\mu + \kp^2/2)}
{\qqp\kkp + a_{\mu\nu}(\qp) -\omega-i\eta} \right\}\;,
\end{eqnarray}
$a_{\mu\nu}(\qp) = \qp^2/2 + \epsilon_\nu - \epsilon_\mu $, $f$ is
the Fermi function at $T=0$, and
$\eta$ is a positive infinitesimal. $\epsilon_\mu$ and
$\varphi_\mu(z)$ are the Kohn-Sham energies and wave function (in LDA)
of the quantum well. For the experimental range of $N_s$,
the system under study has 9 bound levels, only the lowest being
occupied ($N_{\rm occ}=1$).\cite{ullrichvignale}

We consider perturbations
of the form $v_{\rm ext,1}(z,\omega)=  E_0\,z$, corresponding
to monochromatic plane electromagnetic waves of amplitude $E_0$ polarized 
along the $z$-axis, the direction of growth of the quantum well.
Having solved the response equation (\ref{n1bis}), the
photoabsorption cross section is then obtained as
$\sigma(\omega) = -(8\pi \omega/E_0 c)\, {\rm Im} \!
\int\! dz\, z \, n_1(z,\omega)$ and can
be directly compared with data from photoabsorption measurements.
$\sigma(\omega)$ has a peak at the plasmon frequency
$\Omega$ with linewidth (HWHM) $\Gamma$.

In  Fig. \ref{figure1} we plot the dispersions $\Omega(\qp)$ of the
ISB and the intrasubband (or 2D) plasmon in the clean
quantum well ($N_s=1.0\times 10^{11}\:{\rm cm}^{-2}$), calculated within ALDA,
see Eq.  (\ref{fxc_alda}). The imaginary part of the Kohn-Sham response
function $\chi\ks$ determines
the regime of damping by single-particle excitations (Landau damping), as
indicated by the shaded region in Fig. \ref{figure1}. Outside that region,
in particular at small $\qp$, the plasmons are undamped in ALDA, for which
$f_{\rm xc}$ is frequency-independent and real. In reality, however,
the absence of momentum conservation in the $z$-direction and coupling
via Coulomb interaction opens the possibility of plasmon decay
into more complicated excitations, such as multiple electron-hole pairs,
even at $\qp=0$. To take this effect into account, one has to go beyond the
ALDA and include dynamical xc effects.

\begin{figure}
\unitlength1cm
\begin{picture}(5.0,12.5)
\put(-6.3,-8.75){\makebox(5.0,12.5){
\includegraphics{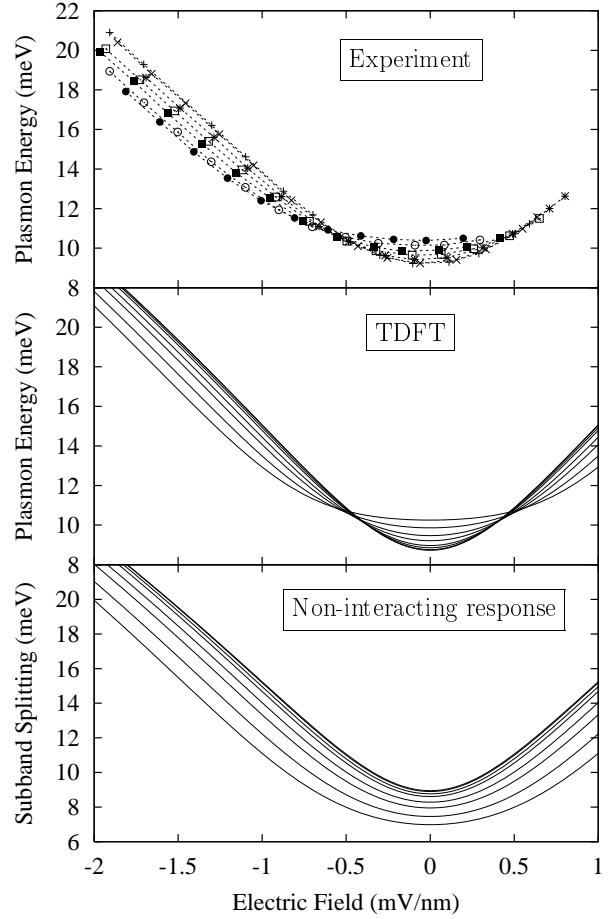}
}}
\end{picture}
\caption{
ISB plasmon frequency $\Omega$, at $\qp=0$, versus electric field $E$. Top:
experimental data from Ref. [23].
Middle: calculated results for the clean quantum well, using TDFT [Eq.
(\ref{n1bis})].  Bottom: bare LDA subband splitting. The individual curves
are associated with different electronic sheet densities ($N_s = 0.05$, 0.1,
0.2, 0.3, 0.5, 0.7, 1.0, and $1.3\times 10^{11}\:{\rm cm}^{-2}$). The lower
$N_s$, the steeper $\Omega(E)$ around $E=0$. }
\label{figure2}
\end{figure}

In Fig. \ref{figure2} we show the electric field dependence of the
ISB plasmon frequencies $\Omega(\qp=0)$ for different values of $N_s$.
In the experimental data, built-in electric fields are
subtracted, so that $\Omega(E)$ exhibits a minimum for $E=0$ and rises
quadratically for small fields. $\Omega(E)$ increases most rapidly
for the smallest $N_s$, since higher electronic densities
tend to screen the external electric field more efficiently.
At the same time, the depolarization shift increases with $N_s$. As a
consequence, the curves of $\Omega(E)$ for different $N_s$
are crossing each other. These features are very well reproduced by theory.
Ignoring dynamical xc effects (i.e. using RPA) induces a $10$\% blueshift of
$\Omega$, which then compares less favorably with experiment. To demonstrate
the importance of including many-body effects in the response equation,
we plot the {\em bare} subband spacings in the bottom panel of
Fig. \ref{figure2}. The results are clearly qualitatively wrong: there is
no crossing of $\Omega(E)$ for different $N_s$, in contradiction to experiment.

\begin{figure}
\unitlength1cm
\begin{picture}(5.0,12.5)
\put(-6.5,-9.4){\makebox(5.0,12.5){
\includegraphics{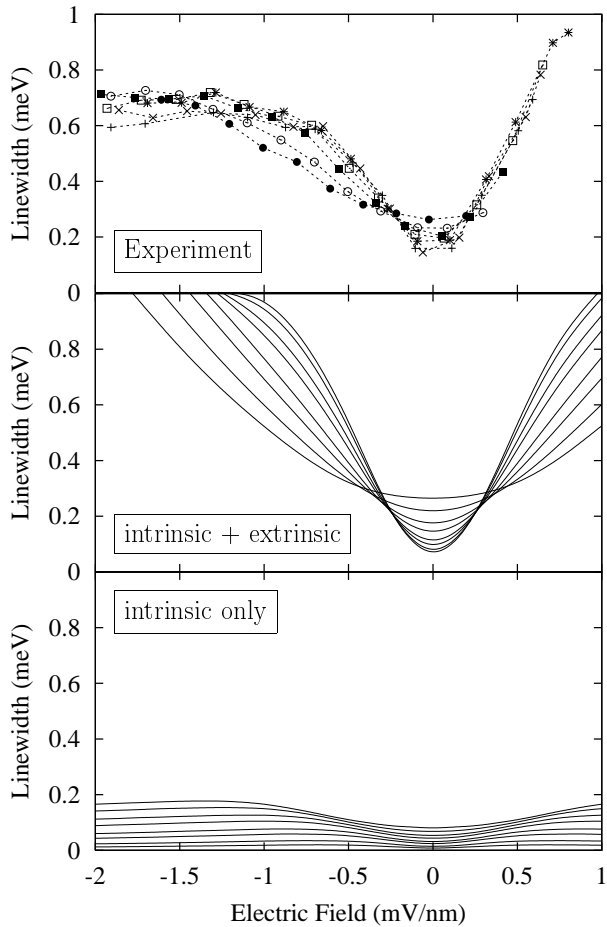}
}}
\end{picture}
\caption{
ISB plasmon linewidth $\Gamma$, at $\qp=0$, versus electric field $E$. Top:
experimental data from Ref. [23]. Middle:
calculated results including extrinsic (impurity and interface roughness)
and intrinsic (electron-electron interaction) damping. Bottom: $\Gamma(E)$
for a clean quantum well (intrinsic damping only).
The individual curves correspond to different electronic sheet densities (see
Fig. \ref{figure2}). At $E=0$, the lowest $N_s$ causes the smallest $\Gamma$.}
\label{figure3}
\end{figure}

Fig. \ref{figure3} shows the
ISB plasmon linewidth $\Gamma(E)$, for different $N_s$.
For small $E$, the experimental data again exhibit a quadratic behavior,
and $\Gamma(E)$ rises faster for smaller $N_s$.
For large negative $E$, $\Gamma$ saturates
around 0.7 meV. For positive $E$ (i.e., pointing in the
direction of sample growth), $\Gamma$ rises
somewhat higher. The asymmetry of $\Gamma(E)$ is likely to be due to
slightly different roughnesses of the interfaces.

The calculated $\Gamma$ for the {\em clean} quantum well
(intrinsic damping through electron-electron interaction only), shown
in the bottom panel of Fig. \ref{figure3},
lies clearly below the experimental values, which is hardly surprising.
However, it can be seen that these purely electronic effects
are far from negligible, at least for $N_s$ not too small, and provide an
intrinsic lower limit to the linewidth of order 0.1 to 0.2 meV for
$N_s \sim 10^{11}\:{\rm cm}^{-2}$.

The middle part of Fig. \ref{figure3} shows $\Gamma(E)$ calculated
including electronic, impurity, and interface roughness damping, using
the combined TCDFT and memory function formalisms outlined above, see Eqs. (\ref{4.A.12}--\ref{4.A.15}).
The results are now in very good agreement with experiment even away
from flat band, as long as $E$ is not too large. We now briefly discuss
the details of how the various contributions to extrinsic damping were
modelled.

The presence of bulk impurities in the quantum well is mainly caused
by segregation of donors from the lower delta-doped layer and diffusion
along $z$ during growth.  We use the functional form
$n_i(z) = 1.33 \: e^{-z/30{\rm nm}} \times 10^{15}\,{\rm cm}^{-3}$,
proposed in Ref. \onlinecite{jon} to explain in-plane mobility data,
for the bulk impurity concentration in Eq. (\ref{memoryfunction_I}).
We also include scattering from the upper delta-doped layer
(remote impurity density $4.8\times 10^{11}\,{\rm cm}^{-2}$).

In Ref. \onlinecite{jon} the in-plane mobility was found to be dominated by
bulk impurity scattering. By contrast, it turns out that neither bulk nor
remote impurities contribute much to the linewidth. The behavior of $\Gamma$
is instead dominated by interface roughness scattering and can
in fact be qualitatively explained by it alone: Via (\ref{memoryfunction_R}),
$\Gamma$ depends on the product of density fluctuations
at the edges, which, for $E=0$, have largest amplitude for highest $N_s$.
For finite $E$, electrons get pushed towards one edge, but
less so for higher densities due to screening of the external field.
$\Gamma(E)$ thus rises more steeply for smaller $N_s$, and the curves cross.

We take a roughness scattering potential of the form
(\ref{roughnesspotential}). The height of the potential step for our quantum
well is $\mu=257.6$ meV. The roughness parameters are chosen as
{$\eta = 64.4$\AA} and $\Delta=4$\AA,
to give the best fit to experiment for the largest $N_s$. Both $\eta$ and
$\Delta$ are in the characteristic range found by lattice imaging techniques.
\cite{microscope}

We also find that including electron-electron scattering does lead to
a significant quantitative improvement for $\Gamma$, in particular for small
$E$.

For $|E|\:\raisebox{-.7ex}{$\stackrel{>}{\sim}$}\: 1\, \rm mV/nm$, the
experimental linewidth saturates. This saturation can be understood
as a negative feedback effect, related to the self-consistency of the
memory functions (\ref{memoryfunction_I}) and (\ref{memoryfunction_R}).
Roughly speaking, the plasmon linewidth comes from the imaginary part of
$M^{L}$, which in turn depends on the imaginary part of $\Phi$.
Broadening of the plasmon resonance means that Im$\Phi$
is peaked around $\Omega$ over some frequency range of width $\Gamma$.
But, due to the constraint of the $f$-sum rule, {\em increasing} $\Gamma$
means that the {\em height} of the peak of Im$\Phi$ must {\em decrease}.
This, in turn, limits the growth the memory function and rapidly saturates
$\Gamma$. Neglect of the self-consistency of the memory function,
as in our calculations, thus means that this saturation effect cannot be fully
captured, as can be seen from Fig. \ref{figure3}.

\section{Conclusion}
\label{secVI}

In this paper, we have dealt with a long-standing
problem in the many-body theory of extended systems: the calculation 
of collective electronic excitations and their associated linewidths in
systems that are both inhomogeneous and weakly disordered, in the sense
that the random potential can be treated as a perturbation. In any real
extended system, collective excitations
are subject to dissipation, causing the associated coherent, plasmon-like 
motion to decay into many individual, incoherent 
degrees of freedom associated with single-particle excitations.
It has been common practice in the literature to describe these processes with phenomenological
assumptions of varying degrees of refinement. This paper, by contrast, presents a new
formalism that allows one to calculate excitation energies and their lifetimes
entirely from first principles. 

Our approach begins with the basic notion that there are two classes of
mechanisms that are responsible for dissipation of collective electronic dynamics.
The first class is {\em intrinsic} scattering, which occurs even in a ``perfect'' material
or device. Here we have in mind primarily systems where phonon scattering is inactive, so
the only source of intrinsic dissipation are electronic many-body effects such as multiple
particle-hole excitations. The second class of dissipation mechanisms is 
{\em extrinsic} in nature, such as scattering off impurities and disorder.

Our treatment of intrinsic scattering relies on TDFT for the linear response.
Fundamental existence theorems guarantee that TDFT describes the linear dynamics 
of interacting many-electron systems in principle exactly, including dissipation
of collective degrees of freedom. In practice, however, the success of a TDFT
approach relies on the approximations used for the linearized xc potential.
The most widely used approximation, the ALDA, has proved to be useful for 
calculating accurate excitation energies, but it produces linewidths that are
strictly zero. Thus, a non-adiabatic description, which includes retardation, is required.
A non-adiabatic dynamical density-functional approach which is local in space but
nonlocal in time has to be formulated replacing the density with the current as basic 
variable (TCDFT). As a result, the linearized xc potential in the 
TCDFT response equation in general acquires a frequency dependence and an imaginary part,
leading to finite linewidths.

To deal with extrinsic scattering, on the other hand, we make use of a powerful
formal technique, the so-called memory function formalism. This approach can be
traced back to the relaxation time approximation, but it replaces the simple
phenomenological relaxation time $\tau$ with the memory function $M({\bf q},\omega)$, which is
defined microscopically as a correlation function between fluctuating random forces.
The memory function formalism is developed in the language of Kubo relaxation
functions, which are, however, intimately connected to the (current)density response 
functions.

The final step then consists in uniting the memory function formalism with
linear response theory in TCDFT. We thus arrive at a new, self-consistent theory that expresses
the response function of an interacting system in the presence of {\em both} intrinsic and
extrinsic damping in terms of the ``clean'' interacting response function
(which contains only intrinsic damping) and the memory function (which accounts only
for extrinsic damping).

We finally applied the theory to describe ISB plasmons in a wide GaAs/Al$_{0.3}$Ga$_{0.7}$As 
quantum well. Using reasonable values for the roughness parameters, we
obtained quantitative agreement with the experimentally measured
linewidth.  But we also found that purely electronic damping due to dynamical exchange
and correlations makes non-negligible contributions to the linewidth, especially
for high electronic densities, where the effect can be as high as a few tens of percents.

A further remarkable outcome of this study is the physical insight that the ISB plasmon
linewidth is primarily controlled by interfacial roughness, and only
weakly affected by the concentration of bulk impurities.
The opposite is true for the in-plane mobility, which is primarily
controlled by bulk impurities.\cite{jon} Thus, the correlation between ISB plasmon
linewidth and in-plane mobility is rather weak, which is physically
understandable since currents are flowing perpendicular to the quantum
well in the former case, and parallel to it in the latter.

\begin{acknowledgments}
This work was supported by NSF grants No. DMR-9706788 and DMR-0074959.
We acknowledge useful discussions with Zhixin Qian, Jon Williams, and Mark Sherwin.
\end{acknowledgments}



\begin{thebibliography}{100}
\bibitem{footnote}
Strictly speaking, the Runge-Gross proof does not apply to
adiabatically switched-on periodic potentials such as the ones employed in
linear response theory.  There are known examples \cite{mearns} of two different
periodic potentials producing the same linear response in a finite  system.
However, such pathologies are expected to disappear \cite{ng1} in extended systems, 
such as the  ones considered in this paper.
\bibitem{footnote2}
It is often stated, not quite accurately, that the GK approximation reduces
to the ALDA in the $\omega=0$ limit. In fact, it turns out that
$\lim_{\omega \to 0}f^h_{\xc}(k=0,\omega) \neq \lim_{k \to 0}f^h_{\xc}
(k,\omega =0) = d^2 \epsilon_{\xc}(n)/dn^2$, as pointed out in Ref.
\onlinecite{contivignale}. The GK interpolation formula for
$f_{\xc}(k=0,\omega)$ does, however, reduce to the ALDA for $\omega \to 0$.
\bibitem{footnote4}
The factor $e/c$ that usually multiplies the interaction $\hat
H_1$ has been absorbed in the vector potential.
\bibitem{HPTfootnote} The exception are harmonically confined
systems, for which the generalized Kohn's theorem
\cite{kohntheorem,dobson1} applies. These systems sustain
collective excitations that are unaffected by intrinsic damping.
\end{thebibliography}
\end{document}